\documentclass{article}

\usepackage{arxiv}

\usepackage[utf8]{inputenc} 
\usepackage[T1]{fontenc}    
\usepackage{hyperref}       
\usepackage{url}            
\usepackage{booktabs}       
\usepackage{amsfonts}       
\usepackage{nicefrac}       
\usepackage{microtype}      
\usepackage{lipsum}		
\usepackage{graphicx}
\usepackage{natbib}
\setcitestyle{numbers,open={[},close={]}}

\usepackage{doi}
\usepackage{subfiles}
\usepackage[section]{placeins}
\usepackage[table,xcdraw]{xcolor}

\usepackage{caption}
\usepackage{subcaption}
\usepackage{multirow}
\usepackage{amsmath}
\usepackage{algorithm}
\usepackage[noend]{algpseudocode}
\makeatletter
\def\algbackskip{\hskip-\ALG@thistlm}
\makeatother
\algnewcommand\algorithmicforeach{\textbf{for each}}
\algdef{S}[FOR]{ForEach}[1]{\algorithmicforeach\ #1\ \algorithmicdo}

\usepackage{xcolor}
\hypersetup{
    colorlinks,
    linkcolor={black!50!black},
    citecolor={black!50!black},
    urlcolor={blue!80!black}
}

\usepackage{titlesec}

\titleclass{\subsubsubsection}{straight}[\subsection]

\newcounter{subsubsubsection}[subsubsection]
\renewcommand\thesubsubsubsection{\thesubsubsection.\arabic{subsubsubsection}}

\titleformat{\subsubsubsection}
  {\normalfont\normalsize\bfseries}{\thesubsubsubsection}{1em}{}
\titlespacing*{\subsubsubsection}
{0pt}{3.25ex plus 1ex minus .2ex}{1.5ex plus .2ex}

\makeatletter
\renewcommand\paragraph{\@startsection{paragraph}{5}{\z@}%
  {3.25ex \@plus1ex \@minus.2ex}%
  {-1em}%
  {\normalfont\normalsize\bfseries}}
\renewcommand\subparagraph{\@startsection{subparagraph}{6}{\parindent}%
  {3.25ex \@plus1ex \@minus .2ex}%
  {-1em}%
  {\normalfont\normalsize\bfseries}}
\def\toclevel@subsubsubsection{4}
\def\toclevel@paragraph{5}
\def\toclevel@paragraph{6}
\def\l@subsubsubsection{\@dottedtocline{4}{7em}{4em}}
\def\l@paragraph{\@dottedtocline{5}{10em}{5em}}
\def\l@subparagraph{\@dottedtocline{6}{14em}{6em}}
\makeatother

\usepackage{acro}
\acsetup{
    first-style = short,
    list/display = used,
    pages/display = first
}

\DeclareAcronym{cvrp}{
  short={CVRP},
  long={Capacitated Vehicle Routing Problem},
  tag = {nomen},
  first-style = short
}

\DeclareAcronym{qubo}{
  short={QUBO},
  long={Quadratic unconstrained binary optimization},
  tag = {nomen},
  first-style = short
}

\DeclareAcronym{cagr}{
  short={CAGR},
  long={Compound annual growth rate},
  tag = {nomen},
  first-style = short
}

\DeclareAcronym{mn}{
  short={Mn},
  long={Million},
  tag = {nomen},
  first-style = short
}

\DeclareAcronym{b2c}{
  short={B2C},
  long={Business to Consumer},
  tag = {nomen},
  first-style = short
}

\DeclareAcronym{roi}{
  short={ROI},
  long={Return on Investment},
  tag = {nomen},
  first-style = short
}

\DeclareAcronym{aqc}{
  short={AQC},
  long={Adiabatic quantum computing},
  tag = {nomen},
  first-style = short
}

\DeclareAcronym{tsp}{
  short={TSP},
  long={Traveling Salesman Problem},
  tag = {nomen},
  first-style = short
}

\DeclareAcronym{vrp}{
  short={VRP},
  long={Vehicle Routing Problem},
  tag = {nomen},
  first-style = short
}

\DeclareAcronym{qpu}{
  short={QPU},
  long={Quantum Processing Unit},
  tag = {nomen},
  first-style = short
}

\DeclareAcronym{rl}{
  short={RL},
  long={Routing Library},
  tag = {nomen},
  first-style = short
}

\DeclareAcronym{gls}{
  short={GLS},
  long={Guided Local Search},
  tag = {nomen},
  first-style = short
}

\DeclareAcronym{nisq}{
  short={NISQ},
  long={Noisy Intermediate-Scale Quantum},
  tag = {nomen},
  first-style = short
}

\title{Adiabatic Quantum Computing for Logistic Transport Optimization}

\author{ {Juan F. Ariño Sales} \\
	ETSISI\\
	Polytechnic University of Madrid\\
	Madrid, ES 28040 \\
	\texttt{juanarinosales@gmail.com} \\
	\And
	{Ra\'ul A. Palacios Araos} \\
	ETSISI\\
	Polytechnic University of Madrid\\
	Madrid, ES 28040 \\
	\texttt{raul.andres.palacios@gmail.com} \\
}

\hypersetup{
pdftitle={Adiabatic Quantum Computing for Logistic Transport Optimization},
pdfsubject={quant-ph, cs.ET}, 
pdfauthor={Juan F. Ariño Sales, Ra\'ul A. Palacios Araos},
pdfkeywords={Quantum Computing, Quantum Annealing, Adiabatic, QUBO, Optimization, VRP, TSP, Logistics, Supply Chain, Last Mile},
}

\begin{document}
\maketitle

\begin{abstract}
	Current world trade is based and supported in a strong and healthy supply chain, where logistics play a key role in producing and providing key assets and goods to keep societies and economies going. Current geopolitical and sanitary challenges faced in the entire world have made even more critical the role of logistics and increased demands for tuning transport function to keep the supply chain up and running. The challenge is only increasing and growing for the future, thus tackling transport optimization provides both business and social value. Optimization problems are ubiquitous and they present a challenge due to its complexity, where they´re typically NP-hard problems. Quantum Computing is a developing field, and the Quantum Annealing approach has proven to be quite effective in its applicability and usefulness to tackle optimization problems. In this work we treat the Vehicle Routing Problem, which is also a variation of a famous optimization problem known as the Traveling Salesman Problem. We aim to tackle the vehicle optimization problem from the last mile logistic scenario application, with a perspective from the classical and quantum approaches, and providing a solution which combines both, also known as hybrid solution. Finally, we provide the results of the analysis and proposal for the consideration of applications in a near term business case scenario.
\end{abstract}

\keywords{Quantum Computing \and Quantum Annealing \and Adiabatic \and QUBO \and Optimization \and VRP \and TSP \and Logistics \and Supply Chain \and Last Mile}

\newpage

\tableofcontents

\newpage
\section{Introduction}
\label{chapter:introduction}
\pagenumbering{arabic}

The logistic industry, understood as the detailed management and implementation of a business operation, may include the research, sourcing, extraction, manufacturing, production, storing, distributing and transportation of goods and assets. The management of the whole workflow is intensive in data processing and analysis. With the very own specific process depending on the type of goods and assets, there is a key function in the value chain in logistic and supply chain, related to the transportation of assets. Thus, Transport Optimization is an important aspect of logistics improvement. From all the various aspects of the supply chain, we focus our work into the applicability of transport optimization for the “Last Mile” scenario.
\\Transport Optimization is the process of finding the best way of moving assets from one place (the source location) to another (destination), and is impacted by many distinct factors, like shipment analysis, transport cost structures, rates, and schedules, cargo, routes, delivery requirements and needs, etc. Combining all distinct factors makes this problem an extraordinarily complex combinatorial problem, which demands high computing power to find viable solutions, for this situation resembles and non-polynomial complexity challenge. Also known as NP-hard problem, Transport Optimization, may be rephrased as finding the optimal value for a transport function, and this is where it becomes a high prospect match for current quantum technology approach, through the Quantum Annealing technique.
\\Throughout this paper, we present the current state of the art in the quantum computing field and dive into the quantum annealing approach to present a viable solution for the Transport Optimization problem. To find the best approach in terms of technology and time-to market applicability, we describe our technique using a hybrid approach, which exploits both classical and quantum techniques to find an optimization model for the Vehicle Routing Problem using Amazon Braket (and the D-Wave quantum annealer). In both cases we provide the results and finally describe our conclusions on the reasons for using a Classical-Quantum approach and recommendations for next steps in a real world use-case scenario. 

\subsection {Objectives}
\label{section:objectives}

The main objective of this work is to generate a series of algorithms which can solve \ac{cvrp} problem instances, the generated algorithms will be compared between each other in order to determine if a quantum or hybrid algorithm is the best option. 
\\A secondary objective is to determine the effects of different hyper-parameters on how the chosen algorithms explore the solution space, the point of this analysis is to determine the best configuration of hyper-parameters in order to better carry out the comparison between classical and quantum algorithms stated in the main objective.
\\A third objective is to explore possible business use cases for \ac{cvrp} solvers and the economical impact these solvers could have on the market.

\subsection{Outline}
\label{section:outline}

This work is organized as follows:

Section \ref{chapter:introduction} provides an overview of the background information necessary to understand this work.
\\Among this required information one can find descriptions of quantum computing, NP-hardness, the state of the quantum computing field and a description of adiabatic quantum computing, in sections \ref{section:quantum_computing}, \ref{section:nphard_definition}, \ref{section:state_qc_field} and \ref{section:adiabatic_quantum_computing} respectively.
\\This section also provides some analysis of the real-world applicability of the problems treated and possible business use cases in subsections \ref{section:transport_optimization} and \ref{section:last_mile_case}, it also includes a more in-depth description of the problem this thesis attempts to solve and the constraints applied to said problem in subsections \ref{subsection:vrp_problem} and \ref{section:Constraints} respectively.
\\Section \ref{chapter:research} provides an overview of the strategies used to solve the \ac{cvrp} problem.
\\Section \ref{chapter:implementation} details the hybrid approach used to solve the \ac{cvrp}, subsections \ref{section:clustering_phase} and \ref{section:routing_phase} describe the 2 distinct phases used in the algorithm. 
\\This section also describes the fully quantum algorithm used to solve the \ac{cvrp} through adiabatic quantum computing in subsection \ref{chapter:qubo_solver}.
\\Section \ref{chapter:analysis} provides an experimental analysis of the aforementioned algorithms, as well as details on how to best compose the hybrid algorithm.
\\The final section \ref{chapter:conclusions} describes the conclusions of this work.
\subsection{Context}
\label{section:context}

\subsubsection{Quantum Computing}
\label{section:quantum_computing}
A computer is a physical device that helps us process information by executing algorithms, which are well defined instructions, with a well scoped description, to process and analyze information through a well-defined processing task. This processing task is translated into a physical task performed in the computing device. 
\\In a classical computer, these tasks use electromagnetic devices that transfer electronic pulses and perform arithmetical operations using logic gates, to control binary signals, which become the information building blocks, known as bits.  

\begin{figure}[!ht]
	\centering
	\includegraphics[width=0.5\columnwidth]{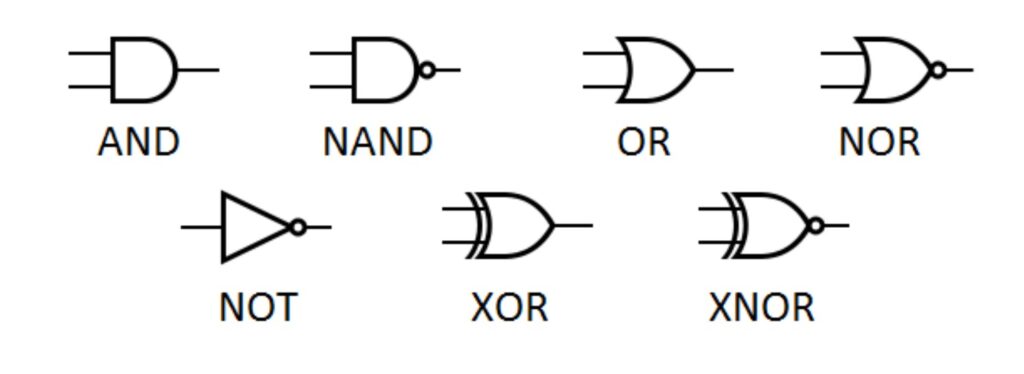}
	\caption{Logic gates \cite{Logic_Gates}}
	\label{fig:Logic_gates}
\end{figure}

In a quantum computer, these tasks use the quantum mechanical principles to perform operation using unitary gates to control a two-state quantum-mechanical system, known as Qubit. can be represented by a linear superposition of its two orthonormal basis states (or basis vectors). These vectors are usually denoted as $ |0\rangle = \begin{bmatrix}
1 \\
0 
\end{bmatrix} $ and $ |1\rangle = \begin{bmatrix}
0 \\
1 
\end{bmatrix} $, which is the Dirac conventional notation (or bra-ket notation), where $ |0\rangle $ and $ |1\rangle $ are known as \emph{ket 0} and \emph{ket 1} respectively, this is called the computational basis, spanning the two-dimensional linear vector (Hilbert) space.

\begin{figure}[!ht]
	\centering
	\includegraphics[width=0.8\columnwidth]{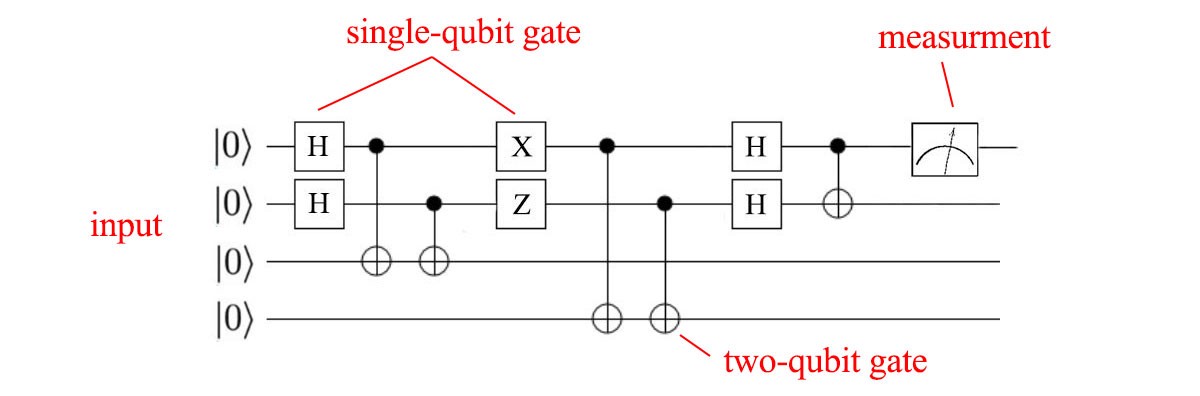}
	\caption{Quantum Circuit \cite{Quantum_Circuit}}
	\label{fig:Quantum_circuit}
\end{figure}

Since a Qubit can be described as a linear combination of $ |0\rangle $ and $ |1\rangle $, the quantum state is said to be described by probabilistic amplitudes of the base vectors describing the Hilbert Space. This results in the Qubit having a wider range of probabilities within its vector space. The probabilistic combinations of these amplitudes must sum up to 100\%, then the following complies: 
\begin{equation}
\label{eq:qubit_state}
|\psi\rangle = \alpha |0\rangle + \beta |1\rangle \hspace{5mm} \textrm{where} \hspace{5mm} |\alpha|² + |\beta|² = 1
\end{equation}

The possible Quantum States for a single Qubit can be visualized using an artifact called the Bloch Sphere, which provides a graphical description of the Qubit probabilistic description using a geometrical polar description of the qubit coordinates, in the Complex numbers plane 

\begin{figure}[!ht]
	\centering
	\includegraphics[width=0.5\columnwidth]{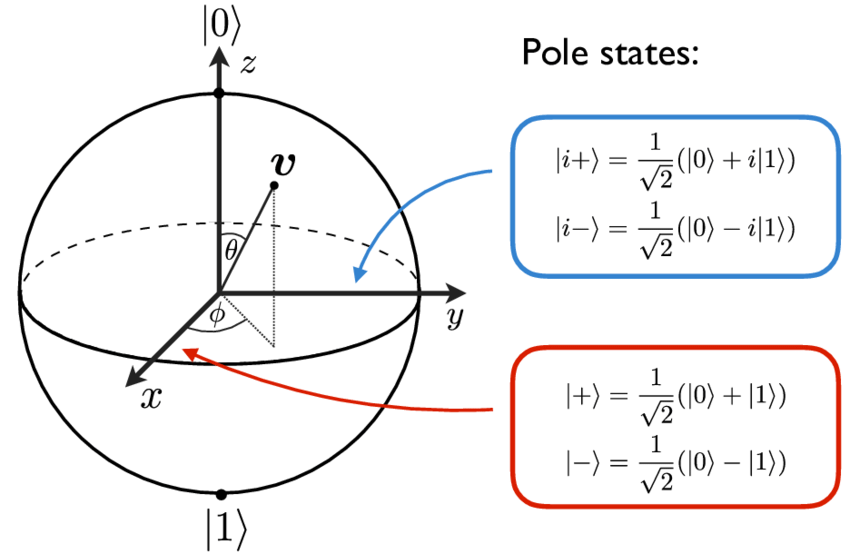}
	\caption{Bloch Sphere \cite{Bloch_Sphere}}
	\label{fig:Bloch_sphere}
\end{figure}

Given this description for the Qubit and its link to a quantum-mechanical system state, it is observable that certain operations can be performed in this quantum state, which then become the building blocks for using quantum mechanics to perform quantum computation through the implementation of quantum circuits using the quantum logic gates approach. These are mathematical operations, becoming unitary operators, described as unitary matrices relative to some basis. 
\\The most common and well know unitary operators are the following: 

\begin{figure}[!ht]
	\centering
	\includegraphics[width=0.5\columnwidth]{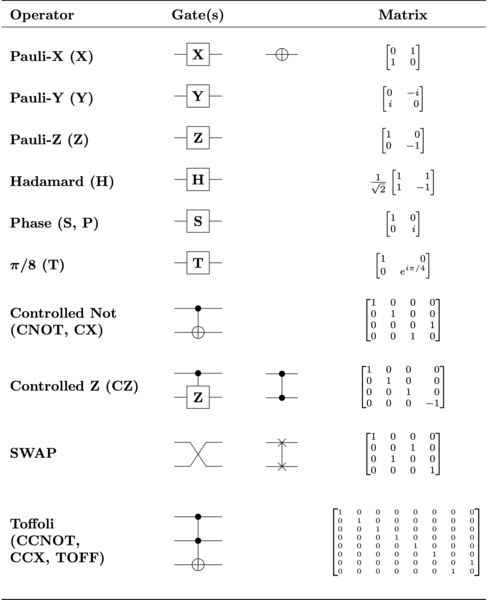}
	\caption{Common quantum logic gates by name, circuit form(s) and matrices \cite{Quantum_Unitary_Operators}}
	\label{fig:Quantum_unitary_operators}
\end{figure}

\subsubsection{State of the Quantum Computing Field}
\label{section:state_qc_field}

Quantum Computing is an emerging technology that is based on the Quantum Mechanic effects for achieving increased parallelism in computation power. Developing Quantum Computing started as the most reasonable way to study the quantum properties of nature. In a quick summary of the history of the development of this field, we can summarize the main milestones: 

\begin{itemize}
  \item Mid 1970´s preliminary attempts to create a Quantum Information Theory.
  \item 1980´s Benioff \& Feynman propose the idea that a computer could operate under the laws of Quantum Mechanics.
  \item In the first half of the 1990´s some important ideas on applicability of quantum computation and initial algorithms arise (Shor´s factorization, Simon´s Oracle later Grove´s algorithm), by the end of the 1990´s the basic ideas of quantum annealing, superconducting circuit and trapped ions transformed into the primary ideas of Quantum Computation applicability, giving rise to the idea of Quantum Computing Industry.
\end{itemize}

\begin{figure}[!ht]
	\centering
	\includegraphics[width=0.7\columnwidth]{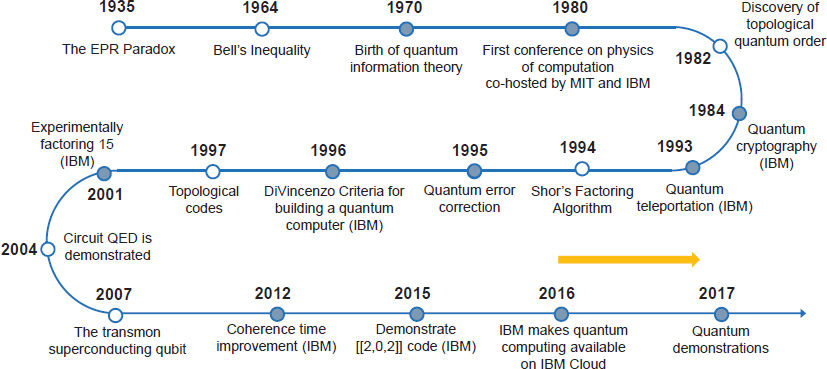}
	\caption{History of quantum computing \cite{QC_History}}
	\label{fig:QC_history}
\end{figure}

Why use quantum computing to tackle complex problems? According to recent research from Boston Consulting Group, 81\% of Fortune 500 companies have use cases to develop within the next 3 years, 31\% of complex problems are abandoned due to the time and resources required to address them.
\\Quantum computing is expected to have US\$ 850 billion market size by 2040 \cite{Building_Quantum_Advantage}. 

\subsubsection{Adiabatic Quantum Computing}
\label{section:adiabatic_quantum_computing}

Adiabatic quantum computation is a form of quantum computing which relies on the adiabatic theorem to do calculations \cite{Quantum_Adiabatic_Evolution}, it is very closely related to quantum annealing, although \ac{aqc} is typically implemented on quantum annealers it has been shown to be polynomially equivalent to circuit model quantum computing \cite{Adiabatic_Quantum_Equivalence}. 
\\The adiabatic theorem states: 
\begin{quote}
    A physical system remains in its instantaneous eigenstate if a given perturbation is acting on it slowly enough and if there is a gap between the eigenvalue and the rest of the Hamiltonian's spectrum \cite{Beweis_des_Adiabatensatzes}.
\end{quote}
In other words, a quantum mechanical system will adapt its functional form when exposed to gradually changing external conditions as long as these conditions vary slowly enough, but it will fail to adapt if the conditions vary too quickly. 

To solve a problem using \ac{aqc} we first need to encode the solution to the problem we are trying to solve as the ground state of a Hamiltonian equation, this will be the final Hamiltonian, next we need to prepare a simple Hamiltonian and initialize it to its ground state, this will be the initial Hamiltonian. 
\\To solve the problem we need to slowly evolve the system from the initial Hamiltonian towards the final one, if this evolution is slow enough the system will remain in the ground state, so at the end the state of the system will describe the desired solution.
\\To apply the Adiabatic Theorem in the Quantum Computing context, the strategy is the following: 
\begin{enumerate}
  \item Design a Hamiltonian whose ground state encodes the solution of an optimization problem.
  \item Prepare the known ground state of a simple Hamiltonian.
  \item Interpolate slowly.
\end{enumerate}

Hamiltonian Equations:

\begin{equation}
\label{eq:final_hamiltonian}
H_1 = \sum h(x) |x\rangle \langle x| \hspace{5mm} \textrm{having} \hspace{5mm} h(x)=n\sum_{u_j u_k \in \bar{E}}x_j x_k - \sum_{u_j \in V} x_j
\end{equation}

\begin{equation}
\label{eq:initial_hamiltonian}
H_0 = I - |s\rangle \langle s| \hspace{5mm} \textrm{having} \hspace{5mm} |s\rangle=\frac{1}{\sqrt{N}}\sum |x\rangle
\end{equation}

\begin{equation}
\label{eq:hamiltonian_evolution}
H = (1-t)H_0 + tH_1
\end{equation}

\subsubsection{NP-Hard Problem Definition}
\label{section:nphard_definition}

The \ac{qubo} model has emerged as an underpinning of the quantum computing area known as quantum annealing and its classical counterpart digital annealing \cite{QUBO_Tutorial}.
\\\acl{qubo} (\ac{qubo}), also known as unconstrained binary quadratic programming, is a combinatorial optimization problem, originally proposed as a metaheuristic model for solving combinatorial optimization \cite{Transverse_Ising}, nowadays has a wide range of applications from finance and economics to machine learning. 
The problem is defined in its most compact way by:

\begin{equation}
\label{eq:qubo}
f_{Q}(x) = \min x^t Q x
\end{equation}

This simple model is notable for embracing a remarkable range of applications in combinatorial optimization. For example, the use of this model for representing and solving optimization problems on graphs, facility locations problems, resource allocation problems, clustering problems, set partitioning problems, various forms of assignment problems, sequencing and ordering problems, and many others have been reported in the literature \cite{QUBO_Survey}. 
\\Let´s review further the problem. A more detailed description of the problem states:
\\Let the function $f_{Q}: \{0,1\}^n \Rightarrow R$ such that: 

\begin{equation}
\label{eq:gen_description_opt_ising}
f_{Q}(x) = \sum_{j=1}^{n}\sum_{k=1}^{j}q_{jk}x_{j}x_{k} + \sum_{j=1}^{n} h_{j} x_{j} + c
\end{equation}

Find a binary vector $x^*$ that is minimal with respect to f among all other binary vectors. This more
generalist approach relates closely to the Ising model, which poses a very similar Hamiltonian.
\\Sometimes, \ac{qubo} is defined as a maximization instead of a minimization problem, which has no effect on the problem's complexity class, as maximizing $f_{Q}(x)$ is the same as minimizing $f_{-Q}(x) = -f_{Q}(x)$

\begin{figure}[!ht]
	\centering
	\includegraphics[width=0.6\columnwidth]{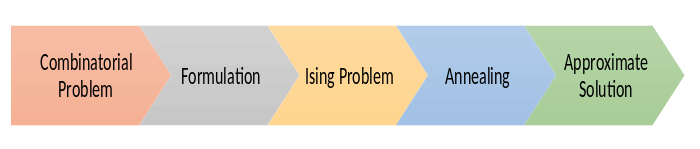}
	\caption{From combinatorial problem towards quantum annealing}
	\label{fig:combinatorial_to_qa}
\end{figure}

\subsection{Business Case}
\label{section:business_case}

\subsubsection{Transport Optimization}
\label{section:transport_optimization}

The Transport Optimization problem, described as the Vehicle Routing Problem is a well-known business scenario where the optimization of logistic processes associated to the loading, programming, clustering and scheduling for vehicles required to follow a path to cover certain destinations under specific constrains, is well measured in terms of business impact. 
\\In the case of the Transport Optimization scenario the range of applicability in logistics is wide and can cover, at the very least, the following possible scenarios: 

\begin{itemize}
  \item Logistics/Material Flow/Assembly Line Optimization 
  \item Grocery Store Logistics
  \item Sustainable Cities: Waste Collection
  \item Public Transportation scheduling 
  \item Traffic Routing \& Traffic Signal Optimization 
  \item Airline Scheduling/Planning, Routing 
  \item Train Platforming Problem
  \item Supply Chain Optimization 
  \item Port Scheduling and Planning 
\end{itemize}

The estimated returns on investments in scenarios for transport optimization range from 40\% to 60\%, since, for instance, Vehicle fleet scheduling is often not optimized effectively, resulting in an unnecessary number of assigned vehicles, lost revenue and waste (e.g., fuel, labor, vehicle wear, etc.), and unnecessary CO2 emissions. This use case is relevant across many industries, including manufacturing, mobility, food delivery, retail, and more. 
\\In a couple of important reference cases, we can consider the piloting and evaluations from Public Transport optimization in Lisbon (Portugal) and New South Wales (Australia). In Lisbon, with car manufacturer Volkswagen; Volkswagen's traffic management system includes two components: passenger number prediction and route optimization using quantum computing. For the pilot project in Lisbon, 26 stops were selected and connected to form four bus links. 
\\Volkswagen experts have developed a quantum algorithm for route optimization between stops. This algorithm calculates the fastest route for each individual bus in the fleet and optimizes it in near real time. In this way, each bus can bypass bottlenecks along the route at an early stage and avoid traffic jams, even before they arise \cite{VW_Traffic_Flow_Optimization}. 
\\In New South Wales, the NSW Government and Transport for NSW have made joint efforts in the application of quantum technologies, for the annual public transportation, passengers take more than 657 million public transport journeys via trains, tubes, buses, ferries and light rail services in NSW and 65 million point-to-point journeys in taxis, car-sharing and rental cars \cite{Quantum_Leap_Travel}. 
\\In a recent study by Boston Consulting Group, the market size for Vehicle Routing Optimization ranges from USD\$ 50B to 100B \cite{Building_Quantum_Advantage}, this is also in line with the high potential scenario for optimization in logistics, from the Study from consulting Company McKinsey \cite{McKinsey_Quantum_Computing} 

\begin{figure}[!ht]
	\centering
	\includegraphics[width=0.6\columnwidth]{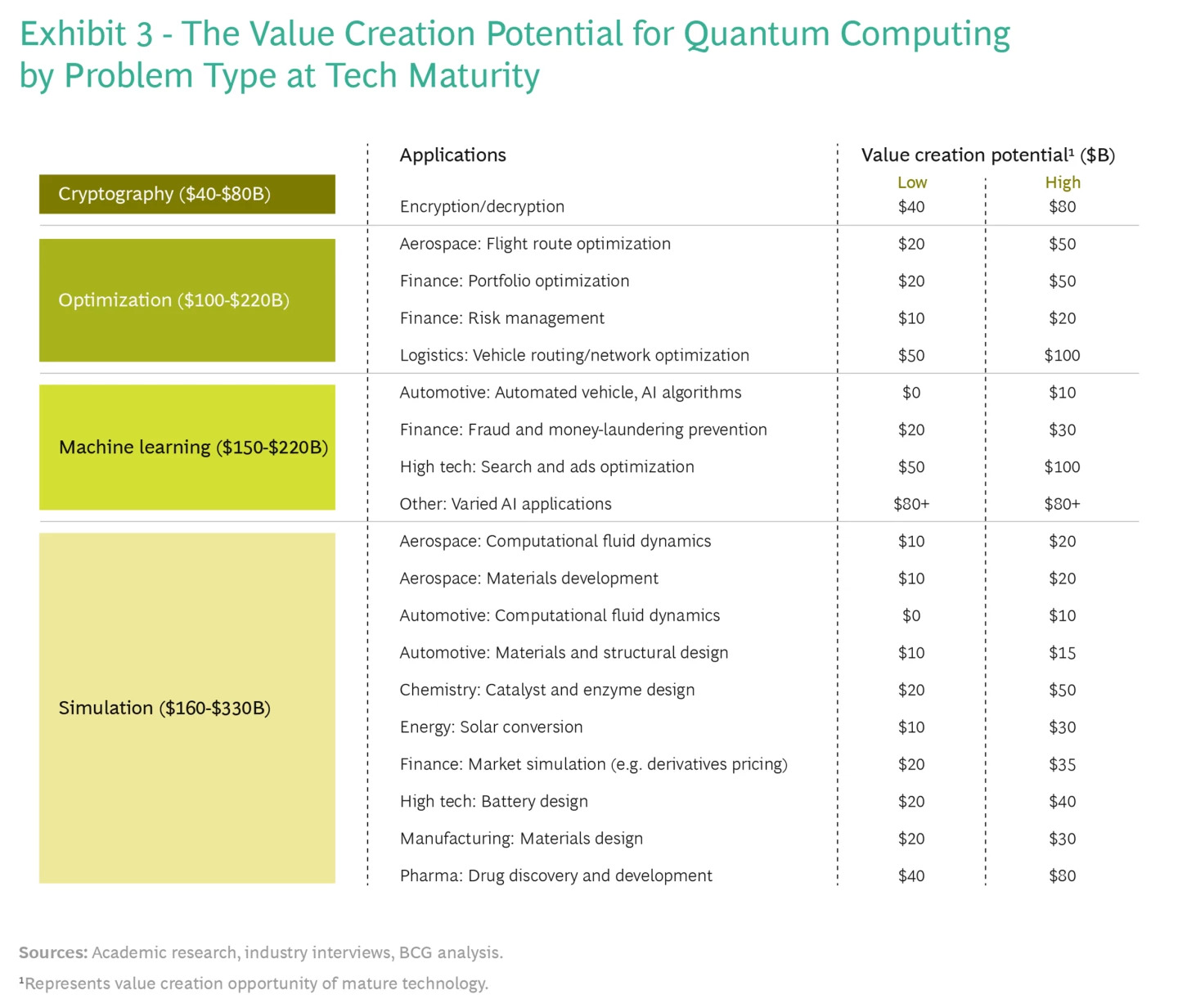}
	\caption{Estimates for the projected value of quantum computing \cite{Building_Quantum_Advantage}}
	\label{fig:Estimates_value_creation}
\end{figure}

\begin{figure}[!ht]
	\centering
	\includegraphics[width=0.7\columnwidth]{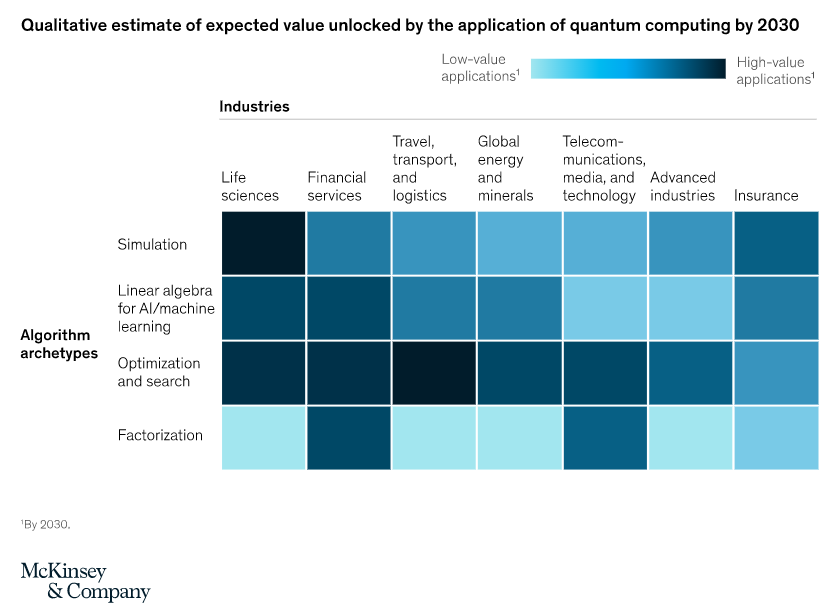}
	\caption{McKinsey estimates Matrix for Value, Industry and Algorithm archetypes \cite{McKinsey_Quantum_Computing}}
	\label{fig:McKinsey_estimate}
\end{figure}

\FloatBarrier

\subsubsection{Last Mile Delivery}
\label{section:last_mile_case}

For our research, we have decided to face the Transport Optimization problem, from the Supply Chain and specifically from the Last Mile delivery challenge, due to its high applicability towards current business challenges. 

\begin{figure}[!ht]
	\centering
	\includegraphics[width=0.8\columnwidth]{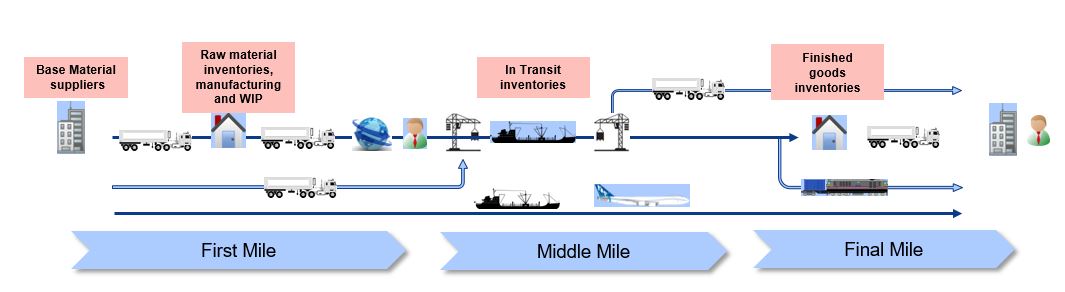}
	\caption{Supply Chain process decoupling \cite{Supply_Chain_WHS}}
	\label{fig:Supply_chain_whs}
\end{figure}

An important part of what is currently known as business state of art is related to understanding the value chain from business and how decoupling business functions may provide agility and scalability to different businesses related to logistics and transportation, looking towards the specialization and optimization of each function, providing and creating value where each business can play the best. Usually, the supply chain can be split into three main stages: the first mile, where the base materials are supplied, arranged, stored, manufactured and/or distributed; the middle mile, where the main transit occurs between the suppliers and the requester, and the last mile, where the finished goods and assets are distributed to its destination to the requester, or customers 
\\The goal of last mile delivery is to transport an item to its recipient in the quickest way possible. This has been driven by the continuously evolving market and demand for convenient customer experience across industries such as e-commerce, food, retail, and many more.
\\There are many crucial elements involved in the last mile delivery process that customers are looking for, namely the speed, accuracy, time efficiency and precision of the product deliveries after reaching their endpoint. 
\\The boom of e-commerce, propelled by the latest geopolitical and sanitary challenges, has made that, this exponential growth and change in the demands from customers for better solutions, is pushing businesses to seek better solutions as a way of responding to social demands and new economic rules. 
\\The last mile delivery market has been steadily growing in the last decade, and the forecasted opportunity follows the same path. The last mile delivery market in Europe is expected to grow from US\$ 677.0 \ac{mn} in 2018 to US\$ 2,491.8 \ac{mn} by the year 2027 with a \ac{cagr} of 16.1\% from the year 2019 to 2027 \cite{Europe_Last_Mile}, while in Spanish speaking Latin America, the level of investment for the last five years is close to US\$ 300 \ac{mn}, leaving Spanish speaking countries like Mexico, Colombia, Chile and Argentina without a leading independent last-mile logistics company, where 60\% of the last-mile delivery market is dominated by small, informal companies. This results in inefficiencies due to a lack of technologies such as route optimization as well as a lack of operating scale.  
\\These issues are quickly becoming more pronounced as e-commerce in Latin America has taken off at a compound annual industry growth rate of 16\% over the past five years \cite{Last_Mile_Latin_America}. In the case of Latam, the biggest e-commerce and retailers have made the last mile logistics, the key value differentiator for growth, leveraging the technology tools and analytical processes, to make investments and plannings in advance \cite{Mercado_Libre_Falabella}. 
\\The situation in Europe is quite similar for the importance of the optimization in the Last Mile transportation, where key factors driving the region’s market growth include rapid industrialization, the growth of the e-commerce sector, and the presence of large and established logistics players. While Germany is a predominant player in the European market, the main segment responsible for its growth is the \acl{b2c} sector. In Spain, the Last Mile market is mostly indexed to the \ac{b2c} sector, which is accountable for over US\$ 40 billion e-commerce market size, here the last mile represents around 40\% of total costs of logistics operations, in a market dominated up to 80\% by small or micro-enterprises \cite{Last_Mile_Logistics_Deloitte}.
\\The fact that there are common components in the last mile market makes the proposal from this paper appealing for a close term application of the technology and solution. In a rough estimate, for a market of USD 27 billion in Spain, with an average of 10\% margin, where the Last Mile may represent something between 30\% to 40\% of the total cost, we aim for a 15 billion market, split in a granular small to micro enterprise sector, which is reported to be close to 2000 \cite{Informe_Sector_Postal}, any 1\% savings in optimization can be worth a very competitive Return on Investment.
\\The fact that there are common components in the last mile market, makes the proposal from this paper appealing for a close term application of the technology and solution. In a rough estimate, for a market of USD 27 billion in Spain, with an average of 10\% margin, where the Last Mile may represent something between 30\% to 40\% of the total cost, we aim for a 15 billion market, split in a granular small to micro enterprise sector, which is reported to be close to 2000 \cite{Informe_Sector_Postal}, any 1\% savings in optimization can prove to be worth a much competitive Return on Investment. 
\\Current investment in the Quantum Computing European Network (in Barcelona) plus the required investment for Human Capital, services and entrepreneurship development, may account for an estimated investment of USD 25 million in 5 years, we summarize the business case as shown in table \ref{fig:Preliminary_ROI_estimations}.
\FloatBarrier
\begin{table}
	\centering
	\begin{tabular}{|c|c|} 
		\hline
		\textbf{Market Size} & \$ 27,000,000,000 \\ 
		\hline
		\textbf{Market Margin} & 10\% \\ 
		\hline
		\textbf{Costs} & 30\% \\ 
		\hline
		\textbf{Num. Operators} & 2000 \\ 
		\hline
		\textbf{Estimated Costs} & \$ 810,000,000 \\ 
		\hline
		\textbf{Est. Unitary Costs} & \$ 405,000 \\ 
		\hline
		\textbf{Gross Inv. Estimate} & \$ 5,000,000 \\ 
		\hline
		\textbf{Estimated Savings} & 1\% \\ 
		\hline
		\textbf{Estimated Savings} & \$ 8,100,000 \\ 
		\hline
		\textbf{\ac{roi}} & 62\% \\ 
		\hline
	\end{tabular}
	\caption{Preliminary Return on Investments Estimations}
	\label{fig:Preliminary_ROI_estimations}
\end{table}

\subsubsection{VRP Problem}
\label{subsection:vrp_problem}

\begin{figure}[!ht]
	\centering
	\includegraphics[width=0.8\columnwidth]{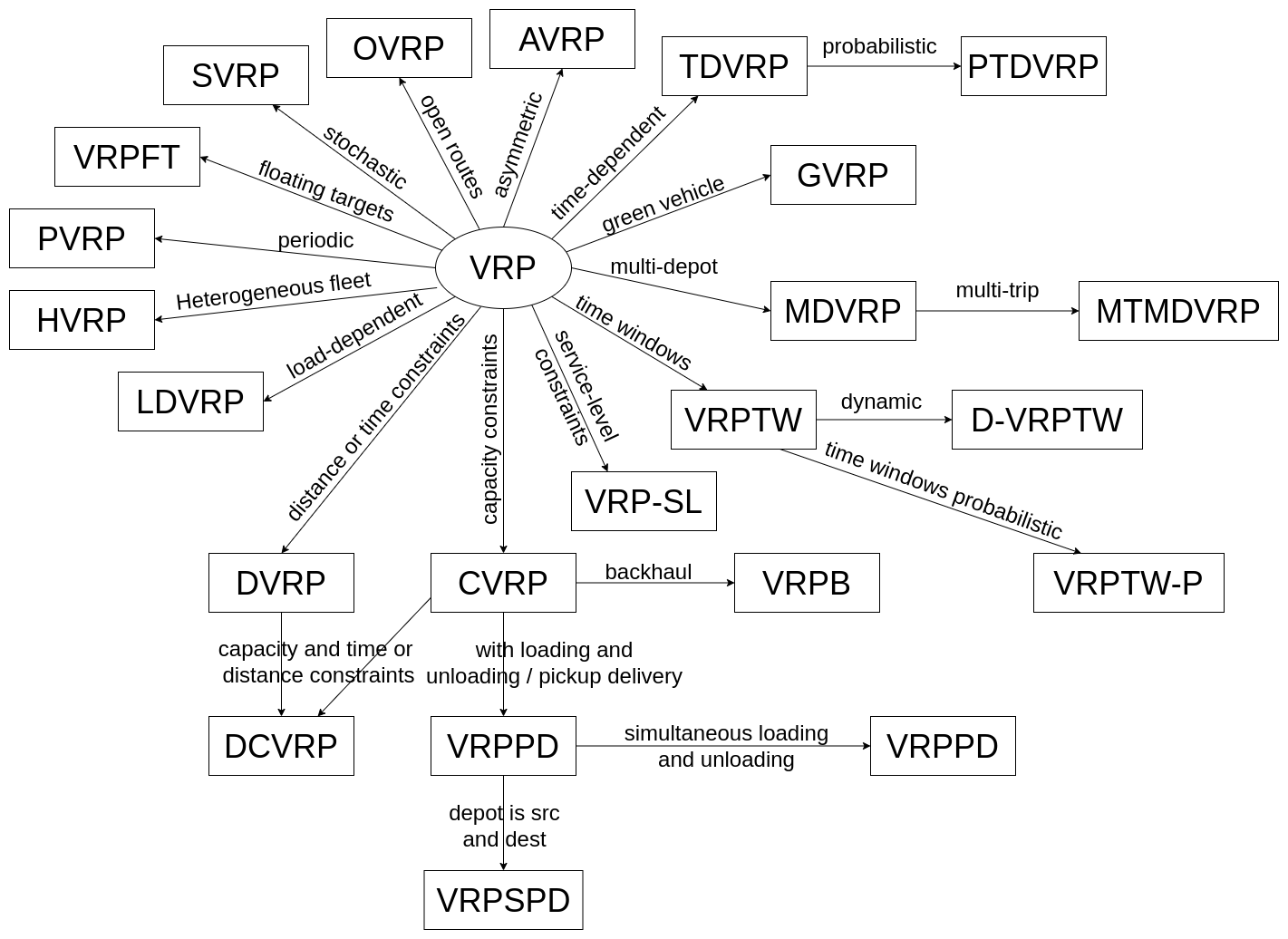}
	\caption{Hierarchy of \ac{vrp} variants \cite{Probabilistic_TD_VRP}}
	\label{fig:VRP_constraints}
\end{figure}

The main problem tackled in this paper is the \acl{vrp} which is an important combinatorial optimization problem where the main goal is to assign an optimal set of routes to deliver some goods from a given origin (depot) to a given set of destinations (customers), it is a generalization of the well-known  \acl{vrp} in which one vehicle must visit several destinations in an optimal fashion. 
\\Both \ac{vrp} and \ac{tsp} are NP-hard problems which makes them exceedingly difficult to solve with classical computing methods since the time taken to solve them grows exponentially with the problem size. Currently the best exact solvers for \ac{vrp} can solve instances with up to 360 customers \cite{VRP_Exact_Solver}, this means that they typically cannot be used to solve real world problems where the number of customers can easily range in the thousands, this number gets even worse once you start introducing constraints such as Time-windows, capacity or others which make it more usable in a real-world scenario. 
\\Heuristic solvers can solve instances with many more customers in a reasonable amount of time, though they are usually designed for a specific problem, and they tend to get stuck in local optimum since they explore a limited solution space, meta-heuristic algorithms solve this problem to a certain extent since they work at a higher level of abstraction which allows them to explore more of the solution space \cite{Genetic_Algorithms_VRP}. Nonetheless both algorithms are hard to use as general solvers as designing the required Heuristics or Meta-heuristics for the specified constraints is usually a challenging task. 
\\The graph below \ref{fig:VRP_constraints} shows some of the more common constraints applied to the \ac{vrp}.

\subsubsection{VRP Constraints}
\label{section:Constraints}

In this paper we tackle the \ac{vrp} with capacity constraints, also known as \ac{cvrp}, with single depot, closed routes and homogeneous fleet constraints, this means that we have a single depot, a vehicle fleet with homogeneous capacity and all routes start at the depot and return to it after servicing the customers, the distances between customers are symmetric but the algorithms used can work with asymmetric distances as well by changing the way the distance matrix is calculated. Some of the conditions imposed are that there must be enough vehicles in the depot to cover all customer demand, each customer is serviced by one vehicle and vehicles cannot return to the depot before servicing all the customers in their assigned route.

\section{Strategy}
\label{chapter:research}

\subsection{Quantum Annealer}
\label{section:quantum_annealer}
\begin{figure}[!ht]
	\centering
	\includegraphics[width=0.5\columnwidth]{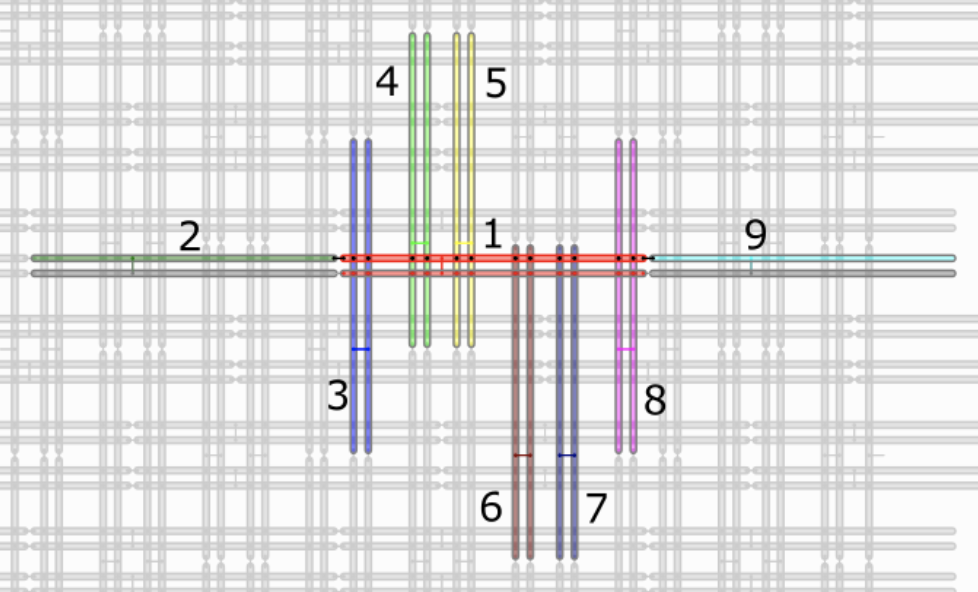}
	\caption{Qubit connectivity inside each tile of the Advantage annealer \cite{dwave_topology}}
	\label{fig:pegasus_graph}
\end{figure}

\begin{figure}[!ht]
	\centering
	\includegraphics[width=0.5\columnwidth]{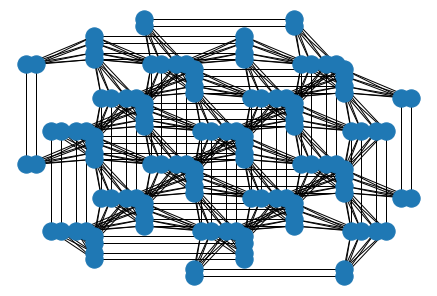}
	\caption{Connectivity structure of a larger Pegasus graph \cite{dwave_braket}}
	\label{fig:pegasus_connectivity}
\end{figure}

The \ac{qpu} used for the experiments performed in chapter \ref{chapter:analysis} is D-Wave's Advantage System 6.1 quantum annealer offered by Amazon Braket, this \ac{qpu} has 5760 qubits and more than 40,000 couplers; The architecture used by this \ac{qpu} is called a Pegasus graph, it is a lattice of 16x16 tiles, each tile is made of a basic unit cell which contains 24 qubits, with each qubit coupled to one similarly aligned qubit in the cell and two others in adjacent cells, each cell is composed by a repeated structure where each qubit is coupled to twelve oppositely aligned, and three similarly aligned, qubits. The structure and connectivity of one tile in the Pegasus graph can be described with figure \ref{fig:pegasus_graph}, image \ref{fig:pegasus_connectivity} shows the connectivity structure of a larger Pegasus graph.

Due to the topology of the \ac{qpu}, the qubits have limited connectivity with each other thus to model a \ac{qubo} problem in the \ac{qpu} we must first find an embedding of the problem onto the Pegasus graph, this is done with a technique called minor embedding.
\\This technique maps one graph onto another graph, it is used to map the graph representing the \ac{qubo} problem onto the Pegasus graph which represents the \ac{qpu} hardware, this is achieved by sacrificing physical qubits according to the connectivity of the \ac{qubo} problem.
\\The minor embedding problem is NP-Hard and for a sufficient number of variables there is no guarantee to find an embedding, with the Advantage \ac{qpu} this limit is around 145 variables, this greatly limits the amount of problems which can be solved with the \ac{qpu}, to overcome this limitation D-Wave offers QBSolv.
\\QBSolv is a hybrid solver that decomposes large \ac{qubo} problems into smaller sub-problems, these smaller sub-problems are then solved using the tabu algorithm or any configured D-Wave solver such as the Simulated Annealing Solver.
\\QBSolv will be discontinued by the end of 2022, it is replaced by dwave-hybrid which is a framework designed to create and test workflows that iterate sets of samples through samplers to solve arbitrary-sized \ac{qubo}s, dwave-hybrid includes a reference example sampler called Kerberos, it finds best samples by running in parallel tabu search, simulated annealing and D-Wave subproblem sampling.

 \begin{figure}[!ht]
	\centering
	\includegraphics[width=0.9\columnwidth]{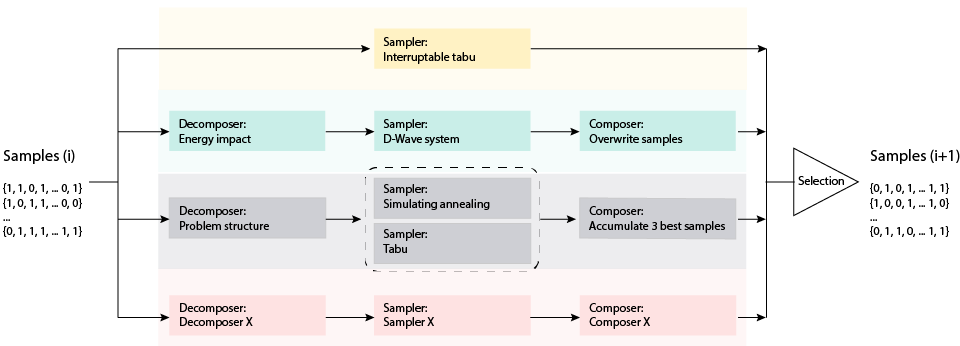}
	\caption{D-Wave hybrid schematic representation with 4 parallel solvers.}
	\label{fig:dwave_hybrid}
\end{figure}

\subsection{Hybrid Algorithm}

The proposed algorithm to solve the \ac{cvrp} problem with the constraints specified in \ref{section:Constraints} is a 2-phase algorithm inspired by \cite{Hybrid_VRP_QA}, this algorithm divides the problem into two separate sub-problems, clustering customers and route optimization, this approach is known as a Cluster-First, Route-Second algorithm, both problems are also NP-Hard but the size of the solution space they need to explore is reduced. 

\begin{figure}[!ht]
	\centering
	\includegraphics[width=0.8\columnwidth]{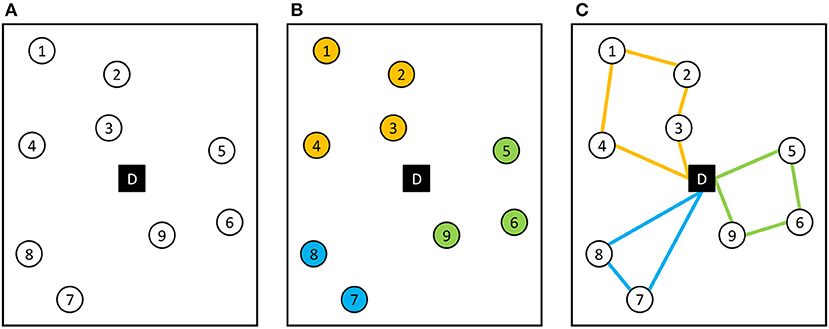}
	\caption{Graphical description of the VRP Problem \cite{Hybrid_VRP_QA}}
	\label{fig:graphical_VRP}
\end{figure}

For each of the 2 phases we consider both a quantum and a classical approach, the results obtained by these differing approaches are compared to select the best one before composing the final hybrid algorithm.  
\\For the clustering phase the algorithms compared are a modified K-Medoids algorithm for the classical approach and a \ac{qubo} formulation of the problem for the quantum approach, these will be explained in more detail in sections \ref{subsection:classical_clustering} and \ref{subsection:quantum_clustering} respectively, the results for the comparison are presented in section \ref{subsubsection:quantum_vs_classical_clustering}. 
\\For the routing phase both presented algorithms attempt to solve the TSP problem, the classical approach is based on combinatorial optimization while the quantum approach models the TSP problem as a \ac{qubo} formulation, a more detailed explanation of the algorithms is given in sections \ref{subsection:classical_routing} for the classical approach and \ref{subsection:quantum_routing} for the quantum approach, the results of the comparison between both approaches are shown in section \ref{subsubsection:quantum_vs_classical_routing}. 

\subsection{QUBO CVRP Solver}

We also propose a fully quantum \ac{qubo} solver, in this solver the \ac{cvrp} problem is modeled with an equation which minimizes the distance traveled by each vehicle, this equation is subject to 6 additional constraints in the form of \ac{qubo} equations to more closely model the \ac{cvrp} constraints explained in section \ref{section:Constraints}.
\\The \ac{qubo} formulation developed for this algorithm is inspired by the formulations presented in \cite{MDVRPC_QA} and \cite{VRP_TW}, a detailed description of the algorithm and its implementation is given in chapter \ref{chapter:qubo_solver}.
\\This algorithm attempts to solve the full \ac{cvrp} problem at once, it is equivalent to running the clustering phase and the routing phase at the same time, thus it has to deal with a much larger solution space than the hybrid algorithm proposed above.

\section{Implementation}
\label{chapter:implementation}

\subsection{Hybrid Algorithm - Clustering Phase}
\label{section:clustering_phase}
\subsubsection{Classical Clustering}
\label{subsection:classical_clustering}

For the classical clustering phase, a K-Medoids algorithm is used to divide the customers into K different clusters, the algorithm is modified to take the total demand of the customers into account such that the distances between customers in the clusters are minimized while the total demand of customers inside each cluster does not surpass the vehicle capacity. 
\\The algorithm is based on the Partitioning Around Medoids algorithm with the added capacity constraint, the steps of the algorithm are the following:

\begin{enumerate}
    \item Select K data points with the highest demand as the medoids.
    \item Determine the clusters by associating each data point to its closest medoid.
    \item Compute the initial cluster costs by adding the distances from every point in each cluster to their medoid, add a penalty cost if the total demand of the cluster exceeds the vehicle capacity.
    \item While the cluster costs decrease and the maximum number of iterations has not been reached:
    \begin{enumerate}
        \item For each medoid $m$ and for each non-medoid data point: $n$
        \begin{enumerate}
            \item Swap $m$ and $o$ and recompute the cluster costs.
            \item If the new cluster cost is higher than the previous one, undo the swap.
        \end{enumerate}
        \item Increase number of iterations.
    \end{enumerate}
    \item Return the clusters.
\end{enumerate}

\begin{algorithm}
    \caption{K-Medoids Clustering}\label{algorithm:kmedoids_clustering}
    \hspace*{\algorithmicindent} \textbf{Input} $M, K, D, Q, I, P$ \\
    \hspace*{\algorithmicindent} \textbf{Output} Clusters Array
    \begin{algorithmic}[1]
    \State $\textit{i} \gets 1$
    \State $\textit{medoids} \gets K \textit{ data points with the highest demand}$
    \BState \emph{compute\_clusters}:
    \State $\textit{clusters} \gets \textit{associate each data point to its closest medoid}$
    \ForEach {$k \in K$}
    \State $\textit{clustercosts}_k \gets \sum_{n \in clusters_k} M_{n,medoids_k}$
    \State $\textbf{if } \textit{clustercosts}_k > Q \textbf{ then } \textit{clustercosts}_k \mathrel{{+}{=}} P$
    \EndFor
    \BState \emph{loop}:
    \If {$\textit{clustercosts} < \textit{prevclustercosts} \And i \leq I$}
    \State $\textit{prevclusters} \gets \textit{clusters}$
    \State $\textit{prevclustercosts} \gets \textit{clustercosts}$
    \ForEach{$m \in \textit{medoids} \And n \not\in medoids$}
    \State $\textit{swap } m \And n$
    \State \textbf{goto} \emph{compute\_clusters}.
    \If {$\textit{clustercosts} > \textit{prevclustercosts}$} undo the swap
    \EndIf
    \EndFor
    \State $i \gets i+1$.
    \State \textbf{goto} \emph{loop}.
    \State \textbf{close};
    \EndIf
    \State $\Return \textit{ clusters}$
    \end{algorithmic}
    \end{algorithm}
    
A pseudocode description of the algorithm is also given in \ref{algorithm:kmedoids_clustering}, in this description $M$ represents the distance matrix with $M_{ij}$ being the distance from node $i$ to node $j$, $K$ is the number of clusters, $Q$ is the vehicle capacity, $D_i$ denotes the demand for customer $i$, $I$ is the maximum number of iterations allowed in the algorithm and $P$ is the penalty cost. 
\\The output is an array of size $|N|$ ($|N|$ is the total number of data points) where the index indicates the customer node and the value which cluster the customer is assigned to.
\\The penalty cost $P$ is added when the cluster load surpasses the vehicle load capacity, it must be carefully chosen to not override the distance cost inside the clusters, but it must be set high enough so that the distance cost does not override the penalty, or else we might obtain clusters with the correct load but increased customer distance or vice versa, clusters with minimal distance between them but with a load higher than the vehicle capacity. 
\\The running time of the algorithm is slightly reduced by precomputing the distance matrix beforehand which allows us to simply add the distances between each point and the medoid instead of recalculating them at every iteration, other techniques such as memoization and vectorization are also used, but the time complexity of the algorithm is still high, $O(K*(N-K)²)$ where $K$ is the number of clusters and $N$ the total number of data points. 
\\The number of clusters chosen is the number of vehicles available in the depot, this way each vehicle will handle one cluster 

\begin{figure}[!ht]
	\centering
	\includegraphics[width=0.5\columnwidth]{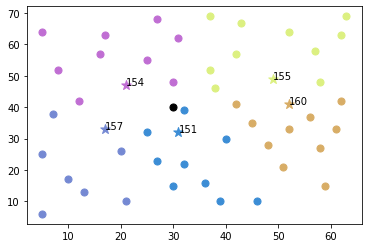}
	\caption{Classical clustering on \cite{Christofides_et_al} data-set, stars represent the chosen medoids and the number is the total demand inside each cluster}
	\label{fig:classical_clustering}
\end{figure}

Other clustering algorithms such as K-means could also be used but K-medoids was chosen due to it being less sensitive to outliers than K-means, one of the main disadvantages of both K-medoids and K-means is that the clusters formed are spherical or convex in shape. Density based clustering algorithms such as DBScan do not have this problem and they typically have a lower time complexity ($O(n²)$ for DBScan), but the number of clusters cannot be specified beforehand, recursive DBScan solves this issue by grouping or splitting the clusters as needed until the specified number of clusters is reached, this is the approach used by other algorithms such as DBScan solver \cite{New_Hybrid_VRP_QA}

\subsubsection{Quantum Clustering}
\label{subsection:quantum_clustering}

For the quantum clustering phase, we propose a \ac{qubo} formulation of the clustering algorithm inspired by \cite{QUBO_Distance_Clustering}, \cite{QUBO_K_Medoids} and \cite{QUBO_Training_ML}, with additional constraints applied to assign each customer to a cluster and consider the total demand of customers in each cluster to ensure that it does not exceed the vehicle capacity.  
\\The proposed formulation is shown below (\ref{eq:qubo_clustering_h}): 

\begin{equation}
\label{eq:qubo_clustering_m}
M = \sum_{k \in K}\sum_{i,j \in I \\ j>i}  dist_{i,j} * x_{i,k} * x_{j,k}
\end{equation}

\begin{equation}
\label{eq:qubo_clustering_c1}
C_{1} = \sum_{k \in K} x_{i,k} = 1 \hspace{5mm}  \forall i \in I 
\end{equation}

\begin{equation}
\label{eq:qubo_clustering_c2}
C_{2} = \sum_{i \in I} d_i * x_{i,k} \leq C \hspace{5mm}  \forall k \in K 
\end{equation}

\begin{equation}
\label{eq:qubo_clustering_h}
H = M + C_1 * M_1 + C2 * M_2 
\end{equation}

$K$ is the total number of clusters while $I$ indicates the customer nodes, $dist_{ij}$ represents the distance matrix between all the possible customer nodes, this matrix is precomputed beforehand. $x_{ik}$ is a binary decision variable which indicates if the customer $i$ is assigned to cluster $k$, $C$ represents the available vehicle capacity and $d_i$ is the demand of customer $i$. 
\\The \ac{qubo} formulation is composed of the main objective function $M$ \ref{eq:qubo_clustering_m} with 2 additional constraints. The formula $M$ tries to find an assignment of customers in clusters such that the total distance between customers in each cluster is minimized, to prevent the formula from not assigning any customer to the clusters, which would obtain a minimum distance of 0, we need to add a penalty for each customer not included in a cluster, this is done through the constraint $C_1$ \ref{eq:qubo_clustering_c1}. 
\\Constraint $C_2$ \ref{eq:qubo_clustering_c2} adds a penalty for each cluster in which the sum of the contained customers demand is greater than the depot vehicle capacity, the formulation is written for a homogeneous vehicle fleet, though extending it to consider a heterogeneous fleet should be trivial, the only change needed is to use a different C value for each cluster. 
\\The multipliers $M_1$ and $M_2$ are used to assign the weight of the corresponding penalty for each of the 2 constraints; these multipliers must be carefully balanced with the main objective function M to find an adequate solution. If $M_1$ is set too high we might obtain a solution where all the customers are assigned to some cluster, but the intra-cluster distance may not be optimal, this typically happens when $C_{1}*M_1$ overpowers the objective function, we might also obtain clusters where the sum of the total demand inside it is higher than the vehicle capacity, this happens when $M_2$ is set too low. To find an optimal solution the weights of the constraints must be balanced with the objective function, the optimal assignment of these parameters may change on a per problem basis.  

\begin{figure}[!ht]
    \centering
    \begin{tabular}{cc}
    
    \includegraphics[width=60mm]{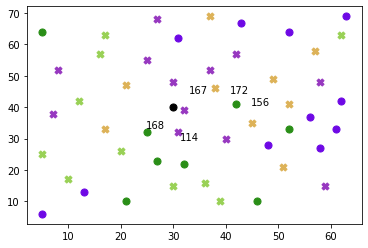}
    &
    \includegraphics[width=60mm]{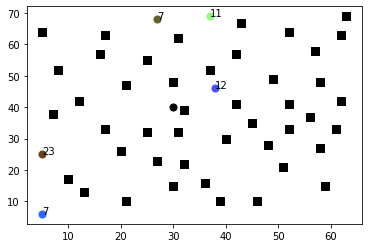}
    \\
    \includegraphics[width=60mm]{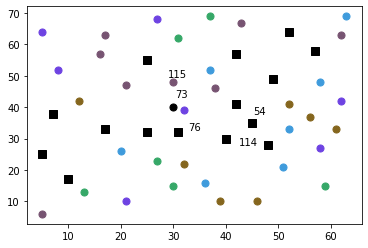}
    &
    \includegraphics[width=60mm]{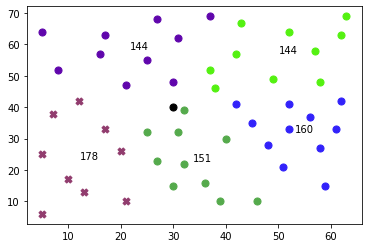}
    \\
    \includegraphics[width=60mm]{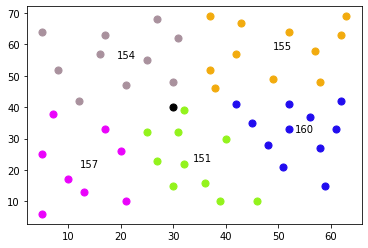}
    \end{tabular}
    \caption{Multiplier effects on the quantum clustering algorithm run on \textit{CMT 01} \cite{Christofides_et_al} data-set}
    \label{fig:parameter_effects}
\end{figure}

In the image \ref{fig:parameter_effects} we can see the effects of the different multipliers, the image on the top left shows the results when the algorithm has $M_1$ set too high, the right top image has $M_1$ set too low, the left middle image has $M_2$ set too high while in the right middle $M_2$ is set too low, the left bottom image shows $M_1$ and $M_2$ well balanced.
\\The X nodes indicate clusters where the cluster demand exceeds the vehicle capacity, black squares indicate nodes without an assigned cluster, the numbers shown in the graphs represent the total demand of the corresponding cluster.

\subsection{Hybrid Algorithm - Routing Phase}
\label{section:routing_phase}
\subsubsection{Classical Routing}
\label{subsection:classical_routing}

For the classical routing phase, the \ac{tsp} is modeled as a combinatorial optimization problem which is then solved using OR-Tools. OR-Tools is an open-source software for solving combinatorial optimization problems developed by Google. 
\\The \ac{tsp} can be modeled inside the OR-Tools framework using the provided \acl{rl} \cite{OR_Tools_RL}. This library allows us to model and solve generic routing problems out of the box, while also allowing for some extensibility to customize the models for specific needs, this greatly simplifies the process of designing an algorithm to solve the \ac{tsp} though it does come with its own set of limitations, namely, there is a limit on the maximum number of nodes that it can handle (32767), a node cannot be visited more than once (except depots which can be starting and ending nodes) and a vehicle cannot transit through a depot.  
\\The \ac{rl} does have other limitations though these can be considered features more than limitations since they are inherent to the solver instead of constraints on the problem, one is the fact that it uses a node-based model, this approach can solve a wide variety of routing problems though there exist certain problems, such as Arc Routing Problems, which are best solved with a different model; The other feature is the fact that the Routing Library returns approximate solutions, since many routing problems are intractable most of the time a solution close to the global optimum is good enough, otherwise it wouldn’t be able to give any solution at all in a reasonable amount of time.  
\\To define a problem using the \ac{rl} we indicate the total number of nodes, the number of vehicles and the depot nodes in the Routing Index Manager, the manager is used to define a Routing Model; We also need to define the distance callback function which will return the distance between any 2 nodes, to save on time we precompute the distance matrix beforehand and return the appropriate value in the callback function, this callback function is added to the Routing Model created earlier and will define the cost of each arc traversed by the algorithm. We also need to define a demand callback function which will return the demand associated with a specific node, the demand callback function is then added to the Routing Model as a dimension variable.
\\With the problem defined in the \ac{rl} we need to specify the search strategy which will be used to find the solution, the search strategy chosen for our classical algorithm is a meta-heuristic strategy called \acl{gls} \cite{OR_Tools_GLS}, though there are many other search strategies available such as Greedy Search, Tabu Search or Simulated Annealing just to name a few, \ac{gls} was chosen since it is especially tailored for the \ac{rl} and has been proven to be particularly successful at solving routing problems \cite{GLS_VRP}.

\begin{figure}[!ht]
	\centering
	\includegraphics[width=0.6\columnwidth]{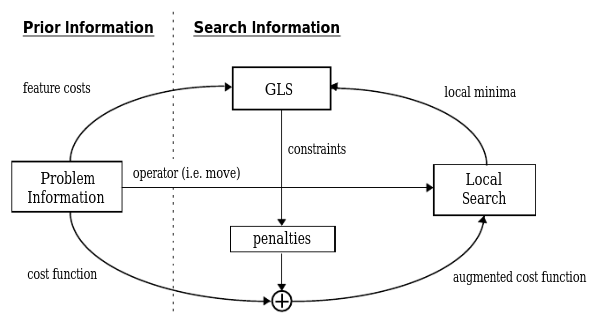}
	\caption{A schematic view of the \ac{gls} approach to combinatorial optimization problems \cite{GLS_Paper}}
	\label{fig:gls_schema}
\end{figure}

\ac{gls} is built on top of a local search algorithm, it gradually adds penalties to certain features of the solutions to help the local search escape from local minima and plateaus, the \ac{gls} implementation in the or-tools framework attempts to minimize the following augmented objective function: 

\begin{equation}
\label{eq:gls_function}
g(x) = \sum_{(i,j)}d_{ij}(x)+\delta \sum_{(i,j)}(I_{ij}(x) \cdot p_{ij} \cdot c_{ij}(x))
\end{equation}

\begin{equation}
\label{eq:gls_penalty}
I_{ij} = 
\left\{ 
  \begin{array}{ c l }
    1 & \quad \textrm{if solution x traverses Arc (i,j)} \\
    0                 & \quad \textrm{otherwise}
  \end{array}
\right.
\end{equation}

Where $ d_{ij} = c_{ij} $ denotes the cost of traversing the Arc from $ i $ to $ j $, $ p_{ij} $ is the penalty assigned to feature $ I $, $ \delta $ is the penalty factor which is used to tune the search to find similar (low $ \delta $) or different (high $ \delta $) solutions.  

\subsubsection{Quantum Routing}
\label{subsection:quantum_routing}

Simulated annealing is used to solve the \ac{tsp} inside each cluster in an adiabatic quantum processor, the \ac{qubo} formulation for the \ac{tsp} used in the algorithm is based on the \ac{qubo} formulation for Hamiltonian cycles implemented in \cite{Ising_Formulations_NP}. The Hamiltonian cycles problem tries to find a path in a given graph which visits every node (besides the starting node) only once and finally returns to the starting node, the \ac{tsp} is a trivial extension of the Hamiltonian cycles problem where the weights of each edge in the cycle are minimized.  
\\The \ac{qubo} formulation for the Hamiltonian cycles problem is given below:

\begin{equation}
\label{eq:qubo_hamilt_c1}
C_1 = \sum_{j \in N+1}(1 - \sum_{i \in N+1}x_{i,j})²
\end{equation}

\begin{equation}
\label{eq:qubo_hamilt_c2}
C_2 = \sum_{i \in N+1}(1 - \sum_{j \in N+1}x_{i,j})²
\end{equation}

\begin{equation}
\label{eq:qubo_hamilt_c3}
C_3 = (1 - x_{0,0})²
\end{equation}

\begin{equation}
\label{eq:qubo_hamilt}
H_A = C_1 + C_2 + C_3
\end{equation}

$ x_{i,j} $ is a binary variable where $ i $ represents the order and $ j $ represents the customers, $ x_{i,j} $  is equal to 1 if the customer with index $ j $ is visited in position $ i $ in the cycle, $ i,j \in {0,…,N} $ where $ N $ is equal to the total number of customers. 
\\The first constraint $C_1$ \ref{eq:qubo_hamilt_c1} ensures that every customer can only appear once in the cycle, the second constraint $C_2$ \ref{eq:qubo_hamilt_c2} ensures that each position in the cycle must be assigned to only one customer, the third constraint $C_3$ \ref{eq:qubo_hamilt_c3} is added so that every cycle starts at customer 0, the resulting \ac{qubo} for the Hamiltonian cycles $H_A$ \ref{eq:qubo_hamilt} is defined by the sum of all three constraints. 
\\To solve the \ac{tsp} problem, we need to add a fourth constraint which will minimize the cost of the edges in the cycle \ref{eq:tsp_hb}.

\begin{equation}
\label{eq:tsp_hb}
H_B = \sum_{h \in N+1} \sum_{i \in N+1, h \neq i} \sum_{j \in N} d_{h,i} x_{j,h} x_{j+1,i} 
\end{equation}

$d$ is the distance matrix which contains the distance between every customer with the depot included as the customer 0, the final \ac{qubo} is the result of adding $H_A$ \ref{eq:qubo_hamilt} and $H_B$ \ref{eq:tsp_hb}. 

\begin{equation}
\label{eq:qubo_tsp}
H = H_A * m_A + H_B * m_B 
\end{equation}

The multipliers $m_A$ and $m_B$ are used to set the penalties for the distinct parts of the equation, they must be carefully chosen so that one constraint does not override the other, if $m_A$ is too high in comparison to $m_B$ the result will be a Hamiltonian Cycle but it might not be the one with the shortest path, if the opposite is chosen, $m_B$ is too high in comparison to $m_A$, the result will be the shortest path but it might not be a Hamiltonian cycle, so there might be some customers left out. 

\begin{figure}[!ht]
	\centering
	\includegraphics[width=0.5\columnwidth]{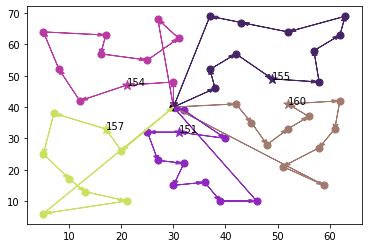}
	\caption{Vehicle routes calculated by the quantum routing algorithm after clustering for the \cite{Christofides_et_al} CMT01 data-set}
	\label{fig:qr_routes}
\end{figure}

\subsection{QUBO CVRP Algorithm}
\label{chapter:qubo_solver}

The \ac{qubo} formulation implemented for this solver is inspired by the formulations presented in \cite{MDVRPC_QA} and \cite{VRP_TW}, it models the whole \ac{cvrp} problem as a \ac{qubo} equation which will be solved with quantum annealing.
\\To model the \ac{cvrp} problem with the constraints specified in section \ref{section:Constraints} as a \ac{qubo} equation we first need to define some parameters, we label the number of available vehicles as $K$ and the total number of customers as $C$, $N$ is the total number of customers plus the starting and ending depots ($N=|C|+2$), the starting depot for the routes is always node 0 and the ending depot is always $|C|+1$, the decision variable for our problem is then defined as follows \ref{eq:vrp_qubo_dec}:

\begin{equation}
\label{eq:vrp_qubo_dec}
x_{ijk} = 
\left\{ 
  \begin{array}{ c l }
    1 & \quad \textrm{if vehicle $k$ travels from node $i$ to node $j$, with ${i,j} \in N$} \\
    0                 & \quad \textrm{otherwise}
  \end{array}
\right.
\end{equation}

Traveling from node $i$ to node $j$ has a cost, this cost is equal to the distance between $i$ and $j$, we label this cost with the variable $c_{ij}$, each customer node also has a certain demand associated with it, we model this demand using the variable $d_i$, each vehicle also has a certain capacity, since we are modeling a homogeneous \ac{cvrp} all vehicles will have the same capacity, this capacity is denoted as $q$.
\\The main \ac{qubo} equation used to minimize the total travel cost is the following \ref{eq:vrp_qubo}. 

\begin{equation}
\label{eq:vrp_qubo}
H = \sum_{k \in K} \sum_{i \in N} \sum_{\substack{j \in N \\ j \neq i}} c_{ij} \cdot x_{ijk}
\end{equation}

This equation is subject to the following constraints:

\begin{equation}
\label{eq:vrp_c1}
C_1 = \sum_{k \in K} \sum_{\substack{j \in N \\ j \neq i}} x_{ijk} = 1 \hspace{5mm} \forall i \in C
\end{equation}

\begin{equation}
\label{eq:vrp_c2}
C_2 = \sum_{j \in N} x_{0jk} = 1 \hspace{5mm} \forall k \in K
\end{equation}

\begin{equation}
\label{eq:vrp_c3}
C_3 = \sum_{i \in N} x_{i,|C|+1,k} = 1 \hspace{5mm} \forall k \in K
\end{equation}

\begin{equation}
\label{eq:vrp_c4}
C_4 = \sum_{\substack{i \in N \\ i \neq h}}x_{ihk} - \sum_{\substack{j \in N \\ j \neq h}} x_{hjk} = 0 \hspace{5mm} \forall h \in C, \forall k \in K 
\end{equation}

\begin{equation}
\label{eq:vrp_c5}
C_5 = \sum_{i \in C}\sum_{\substack{j \in C \\ j \neq i}} d_i \cdot x_{ijk} \leq q_k \hspace{5mm} \forall k \in K 
\end{equation}

\begin{equation}
\label{eq:vrp_c6}
C_6 = \sum_{k \in K}\sum_{i \in S}\sum_{\substack{j \in S \\ j \neq i}} x_{ijk} \leq |S|-1 \hspace{5mm} \forall{\substack{ S \in \mathcal{P}(C) \\ 2 \leq |S| \leq |C|}}
\end{equation}

Constraint $C_1$ \ref{eq:vrp_c1} ensures that each customer is visited only once, constraints $C_2$ \ref{eq:vrp_c2} and $C_3$ \ref{eq:vrp_c3} state that a route must start at depot 0 and end at depot $|C|+1$ respectively, equation $C_4$ \ref{eq:vrp_c4} indicates that after a vehicle arrives at customer $h$ from node $i$ it must leave customer $h$ for node $j$, constraint $C_5$ \ref{eq:vrp_c5} is added to guarantee that the sum of the demand of each customer in the route does not exceed the vehicle capacity $Q$, $q_k = Q \forall k \in K$ since we are modeling a homogeneous \ac{cvrp}.
\\One of the main problems of formulating the \ac{cvrp} as a \ac{qubo} is preventing the formation of impossible routes or closed loops, this happens when a route loops between customers, for example, consider a scenario with one depot, one vehicle and 3 customers, a possible route with closed loops could be $D_1 \rightarrow C_1 \rightarrow D_1$ and $C_2 \rightarrow C_3$ and $C_3 \rightarrow C_2$ , here the route starts and ends at the depot $D_1$ and all customers are visited, however we have a closed loop between customers $C_2$ and $C_3$. To solve this problem we use the constraint $C_6$ \ref{eq:vrp_c6} which is inspired by the Dantzig–Fulkerson–Johnson formulation of the \ac{vrp} \cite{DFJ_tsp}, $\mathcal{P}(C)$ denotes the power set of C. 
\\The constraints presented above need to be reformulated into a \ac{qubo} formulation, the inequality constraints need to be converted to equality constraints, this is achieved using slack variables, which are auxiliary variables whose final value will not be relevant but help the model consider other solutions. The converted constraints are presented below:

\begin{equation}
\label{eq:vrp_qubo_c1}
C_1 = \sum_{i \in C}(1-\sum_{k \in K}\sum_{\substack{j \in N \\ j \neq i}} x_{ijk})²
\end{equation}

\begin{equation}
\label{eq:vrp_qubo_c2}
C_2 = \sum_{k \in K}(1-\sum_{j \in N} x_{0jk})²
\end{equation}

\begin{equation}
\label{eq:vrp_qubo_c3}
C_3 = \sum_{k \in K} (1-\sum_{i \in N} x_{i,|C|+1,k})²
\end{equation}

\begin{equation}
\label{eq:vrp_qubo_c4}
C_4 = \sum_{h \in C}\sum{k \in K} (\sum_{\substack{i \in N \\ i \neq h}}x_{ihk} - \sum_{\substack{j \in N \\ j \neq h}} x_{hjk}) 
\end{equation}

\begin{equation}
\label{eq:vrp_qubo_c5}
C_5 = \sum_{k \in K}((\sum_{i \in C}\sum_{\substack{j \in C \\ j \neq i}} d_i x_{ijk}) + \textit{slack}_{Q}) 
\end{equation}

\begin{equation}
\label{eq:vrp_qubo_c6}
C_6 = \sum_{\substack{S \in \mathcal{P}(C) \\ 2 \leq |S| \leq |C|}} ((\sum_{k \in K}\sum_{i \in S}\sum_{\substack{j \in S \\ j \neq i}} x_{ijk}) + \textit{slack}_{S})
\end{equation}

To convert the inequalities which appear in constraints $C_5$ \ref{eq:vrp_c5} and $C_6$ \ref{eq:vrp_c6} into minimization problems we need to add slack variables, \ref{eq:slack_c5} and \ref{eq:slack_c6} respectively, these are auxiliary binary decision variables whose final value is not relevant but which allow the model some flexibility when considering other solutions.

\begin{equation}
\label{eq:slack_c5}
\textit{slack}_{Q} = \sum_{i \in {0,...,|Q|}} i \cdot s_{i}  \hspace{5mm} 
\end{equation}

\begin{equation}
\label{eq:slack_c6}
\textit{slack}_{S} = \sum_{i \in {0,...,|S|}} i \cdot s_{i}  \hspace{5mm} \forall \substack{ S \in \mathcal{P}(C) \\ 2 \leq |S| \leq |C|}
\end{equation}

\subsection{Requirements}

The \ac{qubo} clustering algorithm, \ac{qubo} routing algorithm and \ac{qubo} solver defined in sections \ref{subsection:quantum_clustering}, \ref{subsection:quantum_routing} and \ref{chapter:qubo_solver} respectively, are implemented in Python 3.9.7 using the PyQUBO package, the \ac{qubo} routing algorithm and \ac{qubo} solver also have different implementations using the Fujitsu Digital Annealer DADK library.
\\The \ac{qubo} equations, generated by the algorithms mentioned above, are solved locally using the Simulated Annealing Sampler (neal) and the decomposing solver QBSolv, both provided by D-Wave. They are also solved on a quantum annealer, D-Wave Advantage System 6.1, using Amazon Braket. 
\\The optimization of the \ac{qubo} equations on real quantum hardware is not as straightforward as with the simulated annealers, depending on the number of variables in the \ac{qubo} problem it might need to be split into smaller sub-problems in order to "fit" inside the \ac{qpu}, a more detailed explanation of this is given in section \ref{section:quantum_annealer}.
\\The K-Medoids clustering algorithm and the classical routing algorithm described in sections \ref{subsection:classical_clustering} and \ref{subsection:classical_routing} are also implemented in Python 3.9.7, the K-Medoids algorithm uses the numpy library to perform some of the computations needed and the classical routing algorithm uses the OR-Tools framework provided by Google.
\\Both classical algorithms (Kmedoids and OR-Routing) as well as the Simulated Annealing Sampler for the \ac{qubo} problems are run on Pop!\_OS 22.04 LTS with a i7-1165G7 Intel CPU and 16 GiB of RAM. 
\\The full code for all the implementations and helper functions can be found at https://github.com/punkyfer/vrpc, the following dependencies are needed to run all the local examples:

\begin{itemize}
    \item dwave-neal==0.5.9
    \item dwave-qbsolv==0.3.4
    \item matplotlib==3.6.0
    \item numpy==1.23.3
    \item ortools==9.4.1874
    \item pandas==1.5.0
    \item pyqubo==1.2.0
    \item scikit-learn==1.1.2
    \item scipy==1.9.1
    \item xmltodict==0.13.0
    \item dadk==2022.4.27
\end{itemize}

\section{Results \& Analysis}
\label{chapter:analysis}

\subsection{Hybrid Algorithm - Clustering Phase}
\label{section:hybrid_algorithm_analysis}
\label{subsection:clustering_phase_analysis}

\subsubsection{Classical Clustering}
\label{subsubsection:kmedoids_clustering_analysis}
\begin{figure}[!htbp]
	\centering
	\includegraphics[width=0.6\columnwidth]{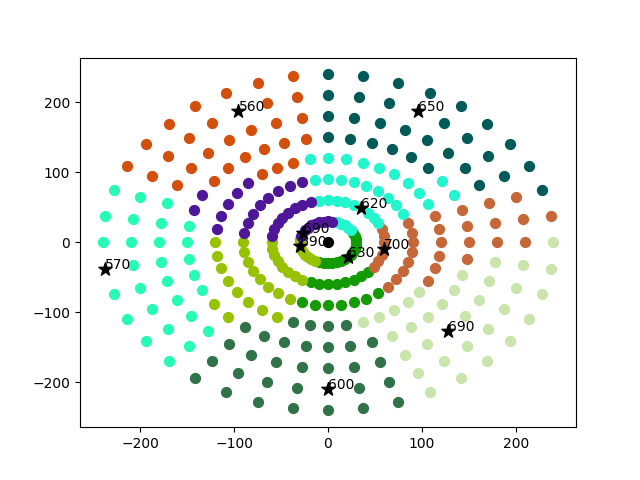}
	\caption{Clusters calculated by the K-Medoids algorithm for the \textit{Golden\_02 data-set} \cite{golden}}
	\label{fig:golden02_kmedoids}
\end{figure}

\begin{figure}[!htbp]
	\centering
	\includegraphics[width=0.6\columnwidth]{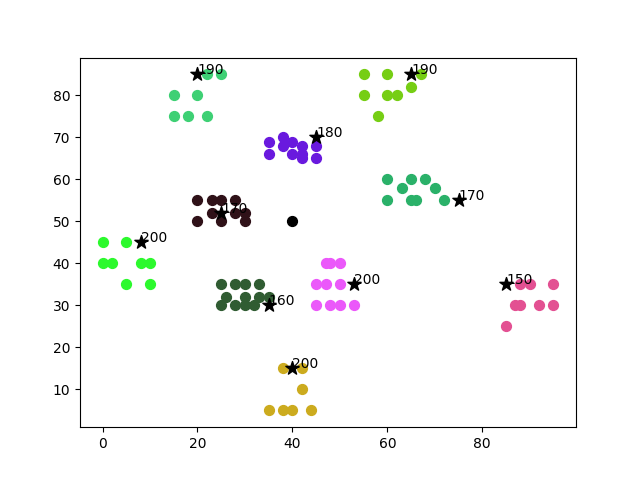}
	\caption{Clusters calculated by the K-Medoids algorithm for the \textit{CMT12} data-set}
	\label{fig:cmt12_kmedoids}
\end{figure}

The K-Medoids clustering algorithm performs a greedy search over the solution space, this means it tends to get "stuck" on local minima, but depending on the structure of the data-set the local minima found is also the global minima. 
\\It performs well on data-sets where the data is structured, this is shown in image \ref{fig:golden02_kmedoids}, or data-sets where the points can be clearly separated into different clusters since in these data-sets there is typically only one minimal solution and thus the greedy search is an adequate method for finding it, we can see this in image \ref{fig:cmt12_kmedoids}.

\begin{figure}[!htbp]
	\centering
	\includegraphics[width=0.6\columnwidth]{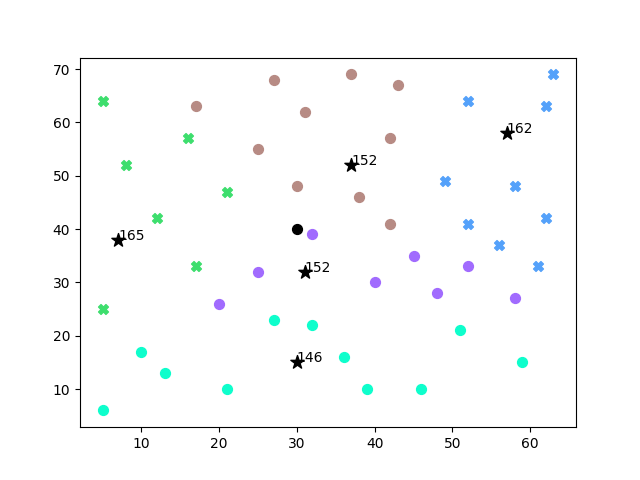}
	\caption{Clusters calculated by the K-Medoids algorithm for the \textit{CMT06} data-set}
	\label{fig:cmt06_kmedoids}
\end{figure}

\begin{figure}[!htbp]
	\centering
	\includegraphics[width=0.6\columnwidth]{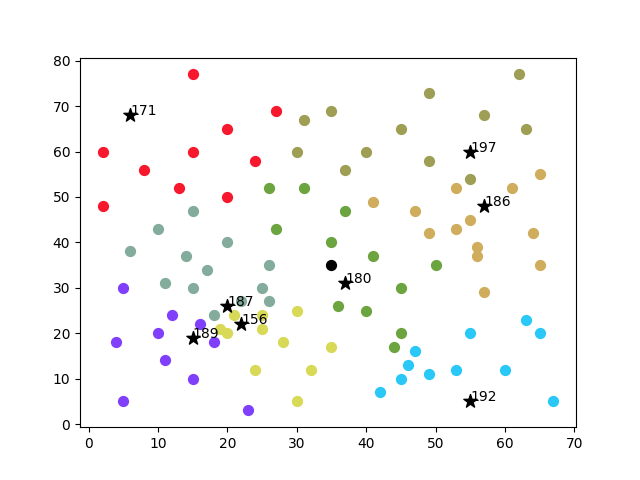}
	\caption{Clusters calculated by the K-Medoids algorithm for the \textit{CMT08} data-set}
	\label{fig:cmt08_kmedoids}
\end{figure}

When the points are more randomly distributed the performance of the K-Medoids algorithm suffers and it struggles to find the global minima, this is due to the fact that these data-sets typically have many local minima where the greedy search can get stuck, we can see this in images \ref{fig:cmt06_kmedoids} and \ref{fig:cmt08_kmedoids}.

\begin{figure}[!htbp]
	\centering
	\includegraphics[width=0.8\columnwidth]{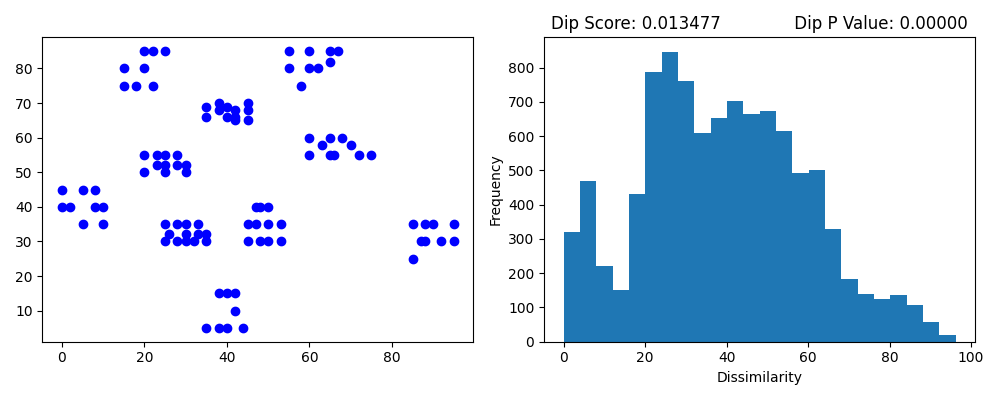}
	\caption{Clusterability score for the \textit{CMT12} data-set}
	\label{fig:cmt12_clusterability}
\end{figure}

\begin{figure}[!htbp]
	\centering
	\includegraphics[width=0.8\columnwidth]{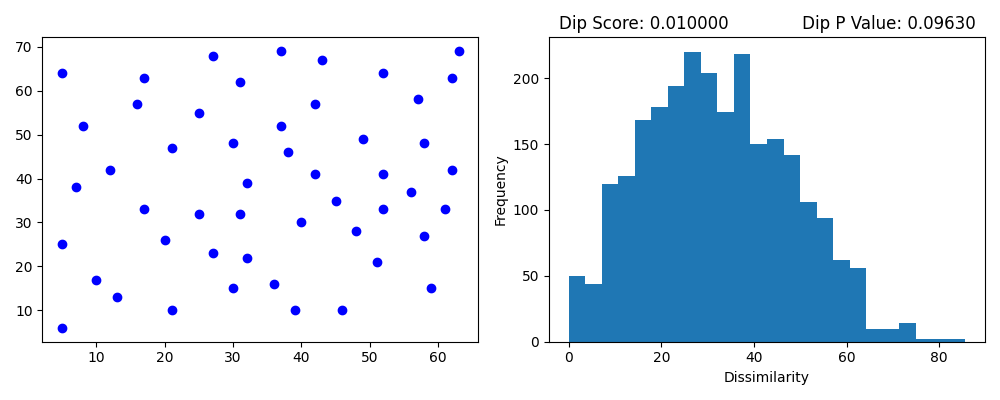}
	\caption{Clusterability score for the \textit{CMT06} data-set}
	\label{fig:cmt06_clusterability}
\end{figure}

\begin{figure}[!htbp]
	\centering
	\includegraphics[width=0.5\columnwidth]{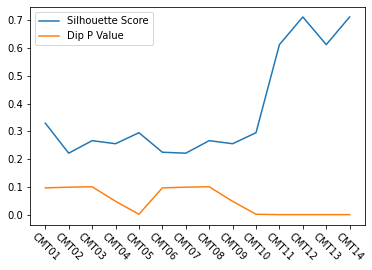}
	\caption{Silhouette score obtained by the K-Medoids algorithm and clusterability score for the \textit{CMT 01-14} data-sets}
	\label{fig:clusterability_silhouette}
\end{figure}

We can observe a relation between the quality of the results obtained by the K-Medoids algorithm and the "clusterability" of the data-set, the better the "clusterability" score of the data, the better the K-Medoids algorithm performs, we can see this in images \ref{fig:cmt12_kmedoids} $\rightarrow$ \ref{fig:cmt12_clusterability} and \ref{fig:cmt06_kmedoids} $\rightarrow$ \ref{fig:cmt06_clusterability}, the relation between the "clusterability" and the silhouette score obtained by the algorithm is shown in figure \ref{fig:clusterability_silhouette}, though it is not a direct relation since there is some variability on the silhouette score obtained by the K-Medoids algorithm.
\\The clusterability score of the data is calculated with the method described in \textit{An Effective and Efficient Approach for Clusterability Evaluation} \cite{clusterability}, first we compute the pairwise distances between all the data points, if the data is "clusterable" the distribution of these distance will be multimodal, with one set representing the small intra-cluster distances and the other the longer distances between different cluster, the multi-modality of the data is checked using Hartigan \& Hartigan's dip test for unimodality \cite{dip_test}, when the resulting P value is low the probability of the distribution being multi-modal is high, and vice-versa.

\FloatBarrier

\subsubsection{Quantum Clustering}
\label{subsubsection:qubo_clustering_analysis}
The \ac{qubo} algorithm used for clustering has 2 multipliers to modify the weights of the different constraints, $M_1$ modifies constraint \ref{eq:qubo_clustering_c1} and $M_2$ modifies constraint \ref{eq:qubo_clustering_c2}, for the experiments we have also added a static multiplier $M_3$ to the distance cost shown in equation \ref{eq:qubo_clustering_m}, the resulting equation is shown below \ref{eq:qubo_clustering_mm}:

\begin{equation}
\label{eq:qubo_clustering_mm}
M = \sum_{k \in K}\sum_{i,j \in I \\ j>i} M_3 * dist_{i,j} * x_{i,k} * x_{j,k}
\end{equation}

\begin{figure}[!htbp]
	\centering
	\includegraphics[width=1\columnwidth]{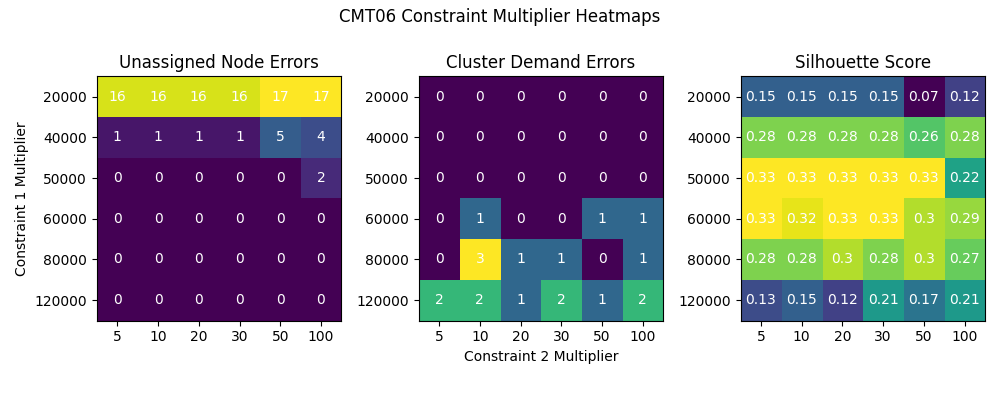}
	\caption{Heat map for the constraint multipliers $M_1$ and $M_2$ for the \textit{CMT06} data-set}
	\label{fig:heatmap_cmt06}
\end{figure}

\begin{figure}[!htbp]
	\centering
	\includegraphics[width=1\columnwidth]{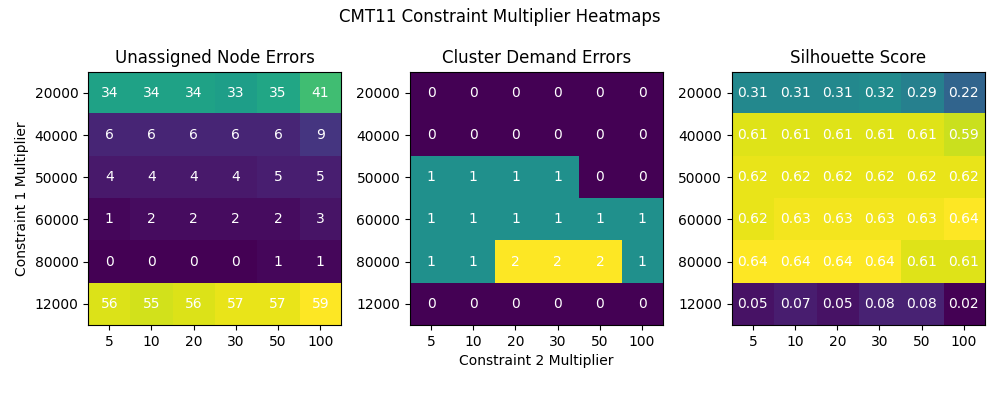}
	\caption{Heat map for the constraint multipliers $M_1$ and $M_2$ for the \textit{CMT11} data-set}
	\label{fig:heatmap_cmt11}
\end{figure}

Multiplier $M_3$ has a static value of 200, to determine the optimal assignment of the other 2 multipliers ($M_1$ and $M_2$) we perform a grid search over some possible values and observe the resulting silhouette score of the clusters generated, as well as the number of unassigned nodes and the number of clusters whose demand surpasses the vehicle capacity, these results are plotted as heat maps, images \ref{fig:heatmap_cmt06} and \ref{fig:heatmap_cmt11} show the heat maps for the data-sets \textit{CMT06} and \textit{CMT11} (\cite{Christofides_et_al}) respectively.

\begin{figure}[!htbp]
	\centering
	\includegraphics[width=1\columnwidth]{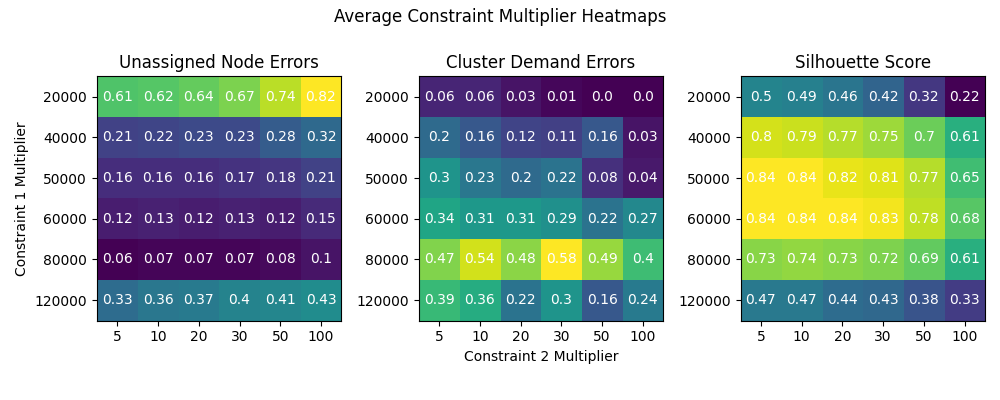}
	\caption{Average heat map for the constraint multipliers $M_1$ and $M_2$}
	\label{fig:heatmap_average}
\end{figure}

We have also plotted a heat map with the averages of all the different normalized results obtained with different data-sets, it is shown in image \ref{fig:heatmap_average}, from this heat map we can observe that the best results obtained by the \ac{qubo} clustering algorithm occur when the $M_1$ multiplier is set to 50,000 and the $M_2$ multiplier to 20, though there is some variability between data-sets these parameters offer good results on average.

\begin{figure}[!htbp]
	\centering
	\includegraphics[width=0.8\columnwidth]{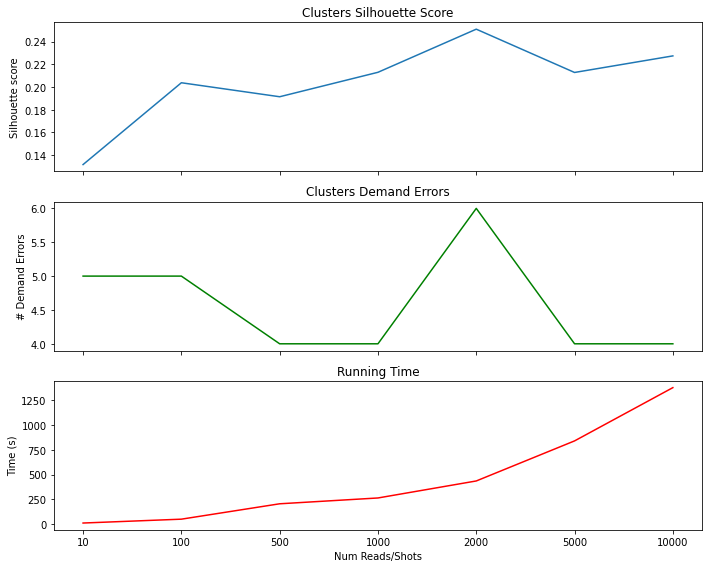}
	\caption{Silhouette score, number of clusters with demand errors and time cost obtained on dataset \textit{CMT02} with different number of reads/shots}
	\label{fig:cmt02_numreads}
\end{figure}

\begin{figure}[!htbp]
	\centering
	\includegraphics[width=0.8\columnwidth]{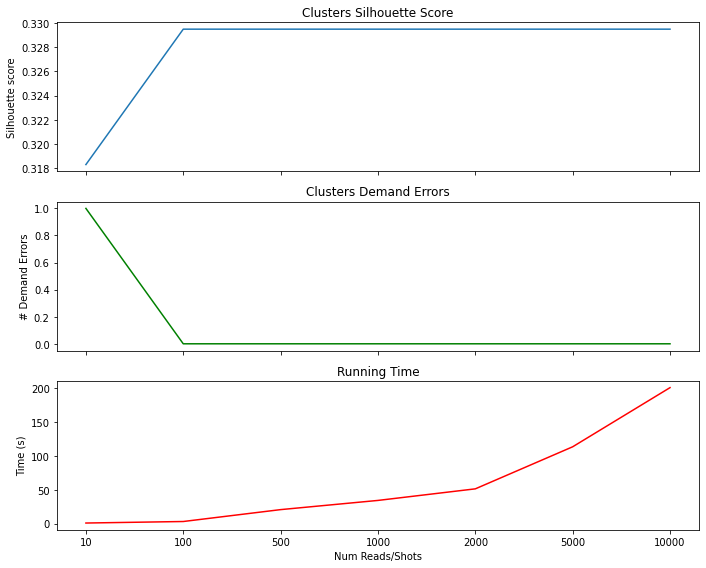}
	\caption{Silhouette score, number of clusters with demand errors and time cost obtained on dataset \textit{CMT06} with different number of reads/shots}
	\label{fig:cmt06_numreads}
\end{figure}

Another parameter we can modify that also affects the quality of the results obtained by the \ac{qubo} clustering algorithm is the number of reads/shots specified in the solver, setting a higher number typically allows the algorithm to explore more of the solution space and return a better solution, we can observe this behaviour in image \ref{fig:cmt02_numreads}, but there comes a point when the algorithm plateaus and increasing the number of reads only increases the computation time without improving the quality of the solution, we can see this happening in image \ref{fig:cmt06_numreads} where the solution stops improving after 100 reads.

Choosing an appropriate value for the number of reads is highly dependent on the number of variables in the \ac{qubo} we are trying to optimize, with higher number of variables requiring a higher number of reads due to the increased size of the solution space, the "clusterability" of the data also plays a role, with more "clusterable" data requiring less reads since the algorithm is able to find the global minima faster.
\\We ran some experiments with the \ac{qubo} clustering algorithm on D-Wave's Advantage System 6.1 provided by Amazon Braket but due to the high number of variables in the \ac{qubo} and the restrictions on the number of qubits available explained in section \ref{section:quantum_annealer} the cost was prohibitively high, table \ref{tab:aws_clustering_costs} shows the costs of clustering some \textit{CMT} data-sets with the \ac{qubo} clustering algorithm using QBSolv() and 100 shots on Amazon Braket. 

\begin{table}[!htbp]
\centering
\resizebox{.7\textwidth}{!}{
\begin{tabular}{|c|c|c|c|c|}
\hline
\textbf{Problem} & \textbf{Nodes} & \textbf{Clusters} & \textbf{QUBO Size} & \textbf{Cost (\$)} \\ \hline
\textbf{CMT01} & 50 & 5 & 515 & 9.57 \\ \hline
\textbf{CMT02} & 75 & 10 & 2140 & 38.28 \\ \hline
\textbf{CMT03} & 100 & 8 & 2392 & 40* \\ \hline
\textbf{CMT04} & 150 & 12 & 4188 & 80* \\ \hline
\textbf{CMT05} & 199 & 16 & 6368 & 120* \\ \hline
\textbf{CMT11} & 120 & 7 & 1540 & 23.37 \\ \hline
\multicolumn{1}{|l}{* Estimated costs} & \multicolumn{1}{l}{} & \multicolumn{1}{l}{} & \multicolumn{1}{l}{} & \multicolumn{1}{l|}{} \\ \hline
\end{tabular}
}
\caption{Costs of running the \ac{qubo} clustering algorithm for some \textit{CMT} data-sets on D-Wave's Advantage System 6.1 provided by Amazon Braket}
\label{tab:aws_clustering_costs}
\end{table}

\FloatBarrier

\subsubsection{Quantum vs Classical Clustering}
\label{subsubsection:quantum_vs_classical_clustering}

\begin{figure}
\centering
\begin{subfigure}{.5\textwidth}
  \centering
  \includegraphics[width=1.1\linewidth]{images/CMT12_kmedoids_clustering.png}
\end{subfigure}%
\begin{subfigure}{.5\textwidth}
  \centering
  \includegraphics[width=1.1\linewidth]{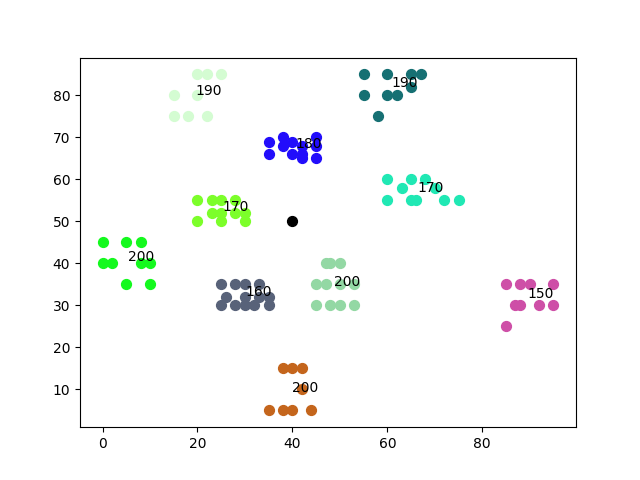}
\end{subfigure}
\caption{Clusters obtained by K-Medoids algorithm (Left) and \ac{qubo} algorithm (Right) for the \textit{CMT12} data-set}
\label{fig:cmt12_qubo_kmedoids}
\end{figure}

From the experiments performed we observe that the 2 clustering functions obtain the same clusters when the data is "clusterable", eg: image \ref{fig:cmt12_qubo_kmedoids}, with the main difference being the running time of each algorithm, where the K-Medoids algorithm outperforms the \ac{qubo} equations (0.36s vs 4.61s in the example shown).

\begin{figure}
\centering
\begin{subfigure}{.5\textwidth}
  \centering
  \includegraphics[width=1.1\linewidth]{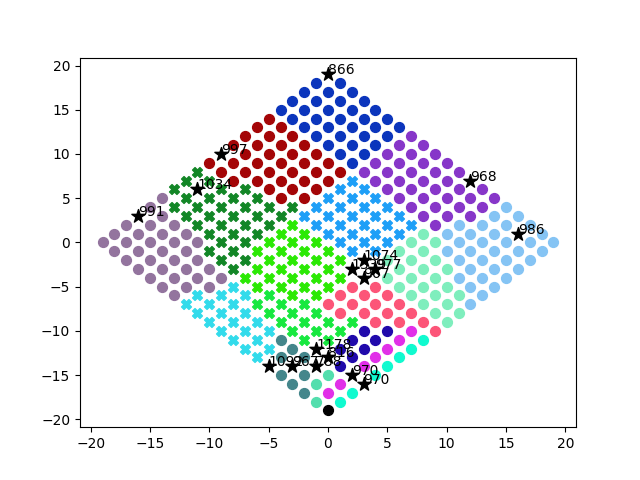}
\end{subfigure}%
\begin{subfigure}{.5\textwidth}
  \centering
  \includegraphics[width=1.1\linewidth]{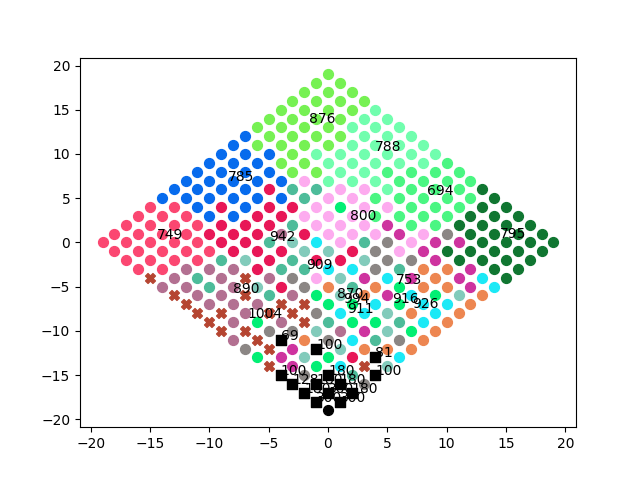}
\end{subfigure}
\caption{Clusters obtained by K-Medoids algorithm (Left) and \ac{qubo} algorithm (Right) for the \textit{Golden\_11} \cite{golden} data-set}
\label{fig:golden11_qubo_kmedoids}
\end{figure}

When the data is structured the K-Medoids algorithm performs better than the \ac{qubo} equations both in quality of the clusters obtained and running time, we can see this in the image \ref{fig:golden11_qubo_kmedoids}, the running time of each algorithm for that example is 61.17 seconds for the K-Medoids algorithm and 1922.55 seconds for the \ac{qubo} equations.

\begin{figure}
\centering
\begin{subfigure}{.5\textwidth}
  \centering
  \includegraphics[width=1.1\linewidth]{images/CMT06_kmedoids_clustering.png}
\end{subfigure}%
\begin{subfigure}{.5\textwidth}
  \centering
  \includegraphics[width=1.1\linewidth]{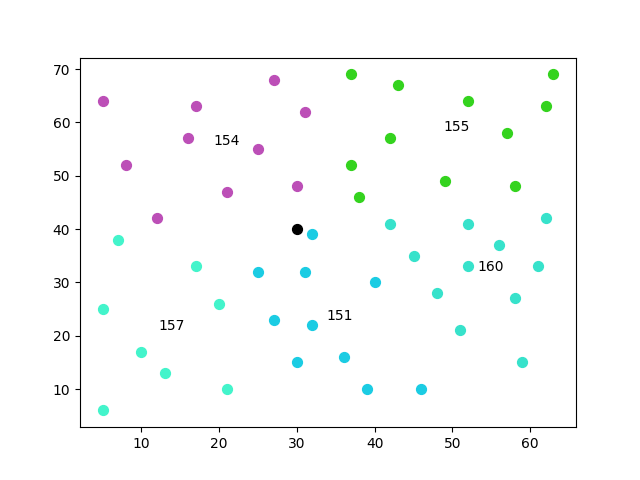}
\end{subfigure}
\caption{Clusters obtained by K-Medoids algorithm (Left) and \ac{qubo} algorithm (Right) for the \textit{CMT06} data-set}
\label{fig:cmt06_qubo_kmedoids}
\end{figure}

However when the data is not structured and it has a low "clusterability" score, the \ac{qubo} algorithm performs better than the K-Medoids algorithm, obtaining a higher silhouette score for the generated clusters and less demand errors in the clusters, this is due to the fact that these solutions typically have many local minima where the K-Medoids algorithm gets stuck while the \ac{qubo} equations are able to surpass these valleys and reach the global minima, we can see this happening in image \ref{fig:cmt06_qubo_kmedoids}.

\begin{figure}
\centering
\begin{subfigure}{.5\textwidth}
  \centering
  \includegraphics[width=1\linewidth]{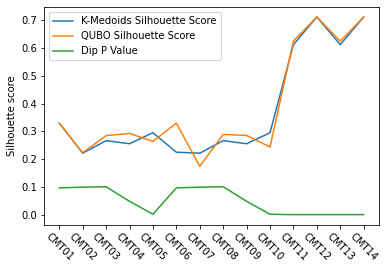}
\end{subfigure}%
\begin{subfigure}{.5\textwidth}
  \centering
  \includegraphics[width=1\linewidth]{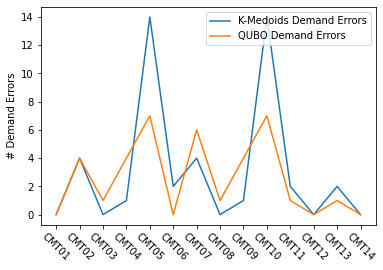}
\end{subfigure}
\caption{Silhouette score (Left) and demand errors (Right) obtained by K-Medoids algorithm (Left) and QUBO algorithm (Right) for the \textit{CMT 01-14} data-sets}
\label{fig:cmt_qubo_kmedoids}
\end{figure}

Image \ref{fig:cmt_qubo_kmedoids} shows the silhouette scores and the number of clusters with demand errors obtained by both algorithms on the CMT data-sets, here we can observe that the \ac{qubo} equations perform slightly better than the K-Medoids algorithm when the data is not clearly "clusterable" (high Dip P value), typically obtaining a higher silhouette score and less number of errors, except in the CMT07 data-set, we can also see that both algorithms obtain the same silhouette score when the data is "clusterable", however the QUBO equations generate clusters with less demand errors than the K-Medoids algorithm.

\begin{figure}[!htbp]
	\centering
	\includegraphics[width=0.6\columnwidth]{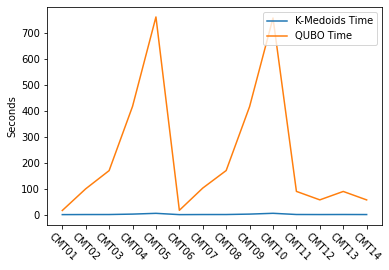}
	\caption{Running time of K-Medoids and \ac{qubo} algorithms on \textit{CMT 01-14} data-sets}
	\label{fig:cmt_qubo_kmedoids_time}
\end{figure}

The running time of the \ac{qubo} equations is much higher than the K-Medoids algorithm, though this is expected since the K-Medoids algorithm is designed to run on a classical processor while the \ac{qubo} equations are designed to be run on a quantum annealer and the experiments are run on the Simulated Annealing Sampler on a classical processor, we can see a comparison of both running times for the CMT data-sets in image \ref{fig:cmt_qubo_kmedoids_time}.

\FloatBarrier

\subsection{Hybrid Algorithm - Routing Phase}
\label{subsection:routing_phase_analysis}

All the experiments for the routing phase are run on the clusters obtained with the K-Medoids algorithm due to the high running time of the \ac{qubo} clustering algorithm.

\subsubsection{Classical Routing}
\label{subsubsection:ortools_clustering_analysis}

\begin{figure}[!htbp]
	\centering
	\includegraphics[width=0.5\columnwidth]{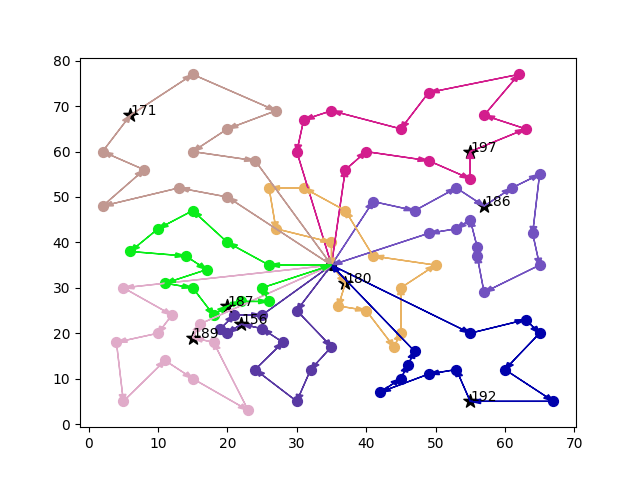}
	\caption{Paths obtained by the OR-Tools routing algorithm for the clusters generated by the K-Medoids algorithm on the \textit{CMT08} data-set}
	\label{fig:cmt08_or_routing}
\end{figure}

\begin{figure}[!htbp]
	\centering
	\includegraphics[width=0.5\columnwidth]{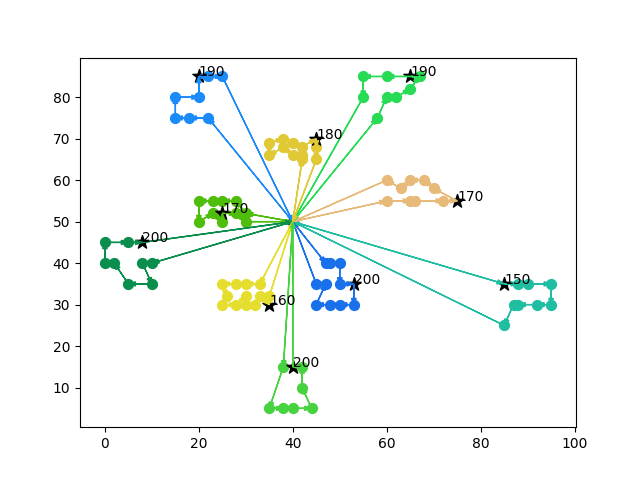}
	\caption{Paths obtained by the OR-Tools routing algorithm for the clusters generated by the K-Medoids algorithm on the \textit{CMT12} data-set}
	\label{fig:cmt12_or_routing}
\end{figure}

The OR-Tools routing algorithm performs a \ac{gls} over the solution space, the \ac{gls} is explained in section \ref{subsection:classical_routing}, but to recap it is a meta-heuristic method on top of a local search algorithm, \ac{gls} builds up penalties while exploring the solution space which help the local search algorithm escape the possible local minima and plateaus in said space, this allows the \ac{gls} algorithm to effectively find the global minima.
\\Images \ref{fig:cmt08_or_routing} and \ref{fig:cmt12_or_routing} show the paths obtained by the OR-Tools routing algorithm when applied to the clusters obtained by the K-Medoids algorithm on the \textit{CMT08} and \textit{CMT12} data-sets, we can see that the algorithm performs well independently of the distribution of the data points.

\begin{figure}
\centering
\begin{subfigure}{.5\textwidth}
  \centering
  \includegraphics[width=1\linewidth]{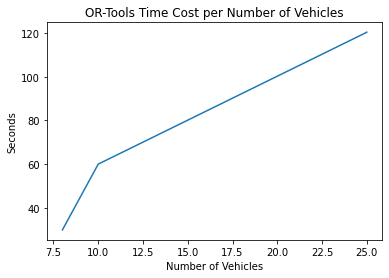}
\end{subfigure}%
\begin{subfigure}{.5\textwidth}
  \centering
  \includegraphics[width=1\linewidth]{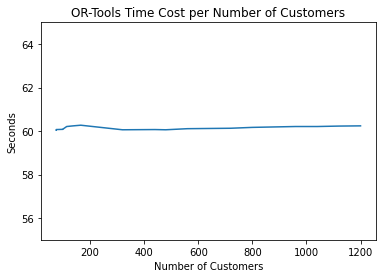}
\end{subfigure}
\caption{Time cost for the OR-Tools routing algorithm for 100 customers and variable vehicles (Left) and for 10 vehicles and variable customers (Right)}
\label{fig:or_tools_timecost}
\end{figure}

We can observe in image \ref{fig:or_tools_timecost} that the time taken to find a solution with the OR-Tools routing algorithm increases with the number of vehicles in the data-set, increasing the number of customers/nodes has no effect on the time cost.

\FloatBarrier

\subsubsection{Quantum Routing}
\label{subsubsection:qubo_routin_analysis}

\begin{figure}[!htbp]
	\centering
	\includegraphics[width=1\columnwidth]{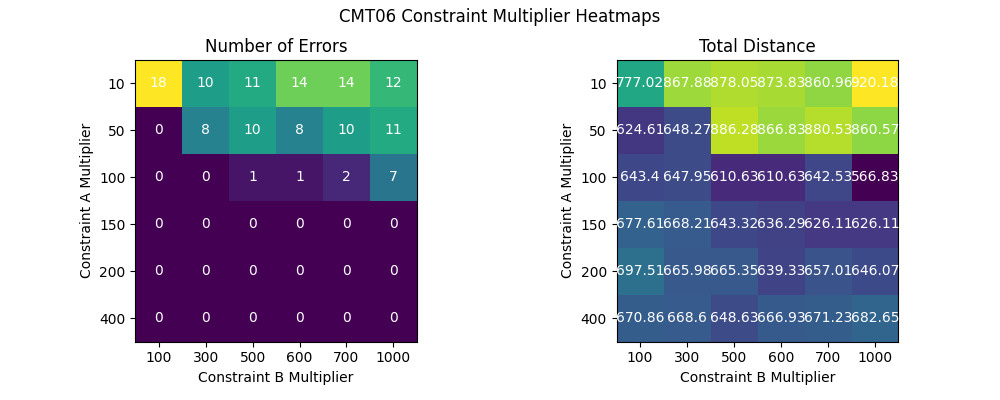}
	\caption{Heat map for the constraint multipliers $m_A$ and $m_B$ for the \textit{CMT06} data-set}
	\label{fig:heatmap_rt_cmt06}
\end{figure}

\begin{figure}[!htbp]
	\centering
	\includegraphics[width=1\columnwidth]{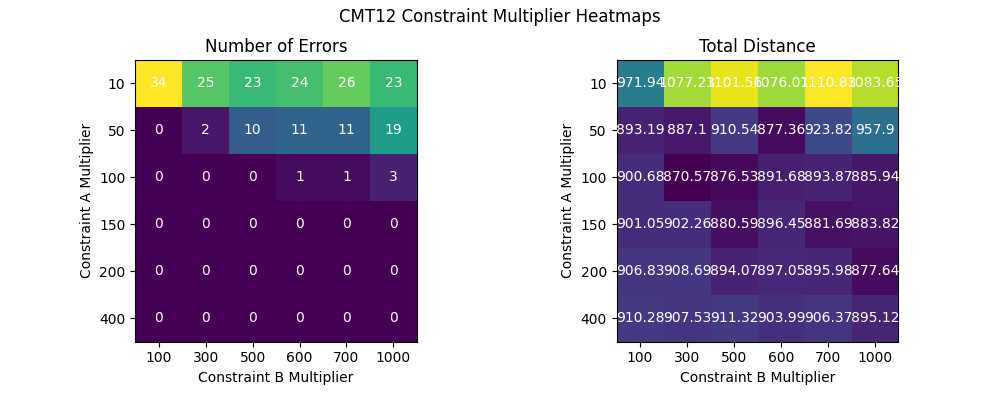}
	\caption{Heat map for the constraint multipliers $m_A$ and $m_B$ for the \textit{CMT12} data-set}
	\label{fig:heatmap_rt_cmt12}
\end{figure}

The \ac{qubo} equations used for routing have 2 multipliers to modify the weights of the different constraints, $m_A$ modifies constraint \ref{eq:qubo_hamilt} and $m_B$ modifies constraint \ref{eq:tsp_hb}, to determine the optimal assignment of these multipliers we perform a grid search over some possible values and observe the sum of the distances for the paths generated, as well as the number of errors in these paths, the results are plotted as heat maps, images \ref{fig:heatmap_rt_cmt06} and \ref{fig:heatmap_rt_cmt12} show the heat maps for the data-sets \textit{CMT06} and \textit{CMT12} (\cite{Christofides_et_al}) respectively.

\begin{figure}[!htbp]
	\centering
	\includegraphics[width=1\columnwidth]{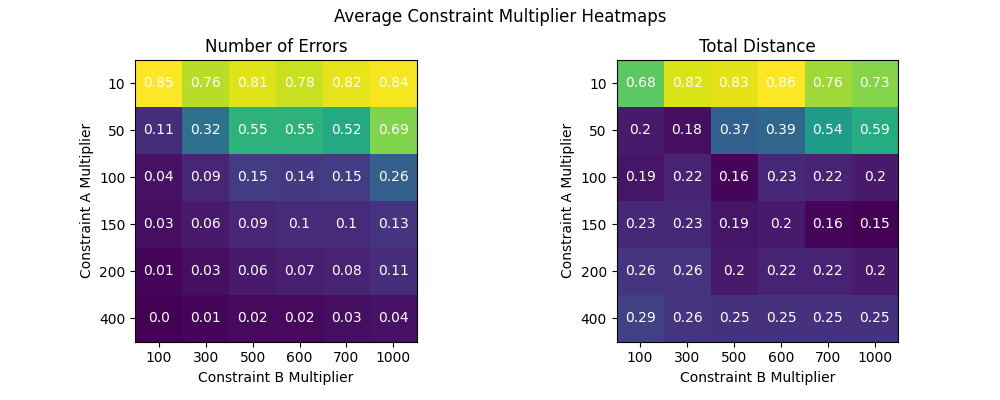}
	\caption{Average heat map for the constraint multipliers $m_A$ and $m_B$}
	\label{fig:heatmap_rt_avg}
\end{figure}

A heat map with the averages of all the normalized results obtained with different data-sets can be seen in image \ref{fig:heatmap_rt_avg}, in this heat map we can  see that the lowest distance, with the least number of errors, is obtained when the $m_A$ multiplier is set to 150 and the $m_B$ multiplier to 700, we can slightly reduce the distance further by setting $m_B$ to 1,000 but the increase in the number of errors is not worth it.

\begin{figure}[!htbp]
	\centering
	\includegraphics[width=0.8\columnwidth]{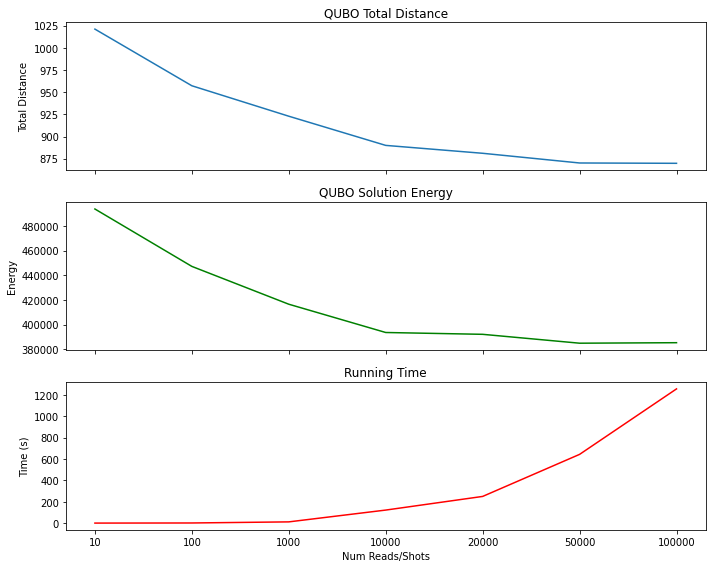}
	\caption{Total distance, \ac{qubo} solution energy and time cost obtained on dataset \textit{CMT12} with different number of reads/shots}
	\label{fig:cmt12_rt_numreads}
\end{figure}

\begin{figure}[!htbp]
	\centering
	\includegraphics[width=0.8\columnwidth]{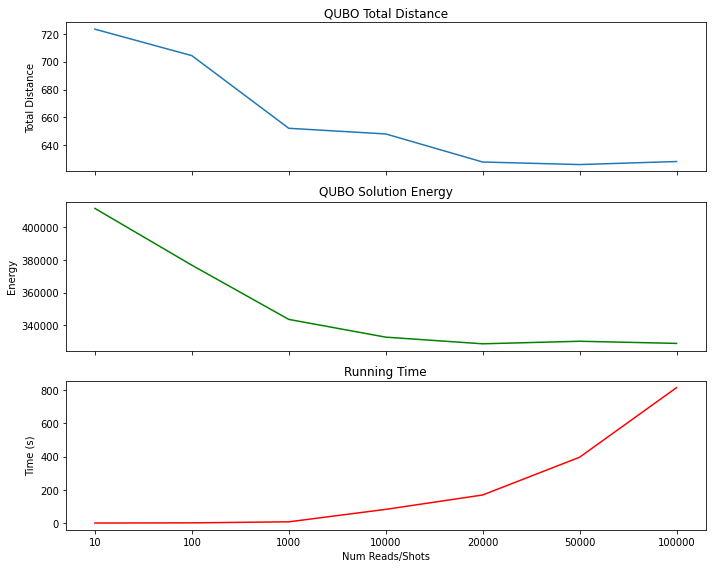}
	\caption{Total distance, \ac{qubo} solution energy and time cost obtained on dataset \textit{CMT06} with different number of reads/shots}
	\label{fig:cmt06_rt_numreads}
\end{figure}

The other parameter of the \ac{qubo} routing algorithm which affects the quality of the results obtained is the number of reads/shots performed by the solver, usually a higher number of reads returns a better solution while increasing the computation time since the algorithm is allowed to explore more of the solution space, we can see this in image \ref{fig:cmt12_rt_numreads}, but depending on the size of the \ac{qubo} increasing the number of reads might not offer better solutions while still increasing the computational time, such as in image \ref{fig:cmt06_rt_numreads}, where the quality of the solution plateaus after 20,000 iterations.

The number of reads/shots necessary to find a good solution is highly dependent on the size and complexity of the problem, it must be carefully chosen to not waste computing time, especially when the solver is a real quantum annealer since it has an increased economic cost versus a traditional classic solver.
\\We ran some experiments with the \ac{qubo} routing algorithm on D-Wave's Advantage System 6.1 provided by Amazon Braket but due to the high number of problems generated the cost is high, one advantage of running the \ac{qubo} routing algorithm after the clustering algorithm is that the problems fit in the quantum annealer without using QBSolv(), table \ref{tab:aws_routing_costs} shows the costs of running the \ac{qubo} routing algorithm on the clusters generated by the K-Medoids algorithm from the \textit{CMT} data-sets, using 2000 shots on Amazon Braket. 

\begin{table}[!htbp]
\centering

\begin{tabular}{|c|c|c|c|}
\hline
\textbf{Problem} & \textbf{Nodes} & \textbf{Clusters} & \textbf{Cost (\$)} \\ \hline
\textbf{CMT01} & 50 & 5 & 6.8 \\ \hline
\textbf{CMT02} & 75 & 10 & 5.44 \\ \hline
\textbf{CMT03} & 100 & 8 & 8.84 \\ \hline
\textbf{CMT04} & 150 & 12 & 8.84 \\ \hline
\textbf{CMT05} & 199 & 16 & 8.84 \\ \hline
\textbf{CMT11} & 120 & 7 & 12.24 \\ \hline
\end{tabular}%

\caption{Costs of running the \ac{qubo} routing algorithm for some \textit{CMT} data-sets on D-Wave's Advantage System 6.1 provided by Amazon Braket}
\label{tab:aws_routing_costs}
\end{table}

\FloatBarrier

\subsubsection{Quantum vs Classical Routing}
\label{subsubsection:quantum_vs_classical_routing}

\begin{figure}[!htbp]
	\centering
	\includegraphics[width=0.5\columnwidth]{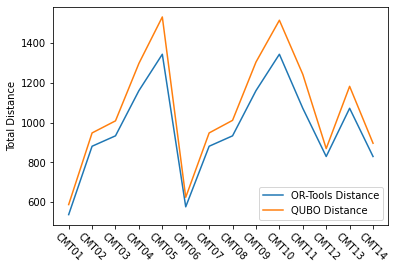}
	\caption{Total distance obtained with OR-Tools and \ac{qubo} routing algorithms on \textit{CMT 01-14} data-sets}
	\label{fig:routing_distance}
\end{figure}

From the experiments performed we can observe that the OR-Tools routing algorithm performs better than the \ac{qubo} equations, obtaining paths with a lower total distance in all the data-sets tested, we can see the total distance of the paths obtained by each algorithm on some of the data-sets in image \ref{fig:routing_distance}, it is clear that the OR-Tools routing algorithm performs better, though the \ac{qubo} equations do come close on some data-sets the OR-Tools algorithm still obtains slightly better results.

\begin{figure}
\centering
\begin{subfigure}{.5\textwidth}
  \centering
  \includegraphics[width=1\linewidth]{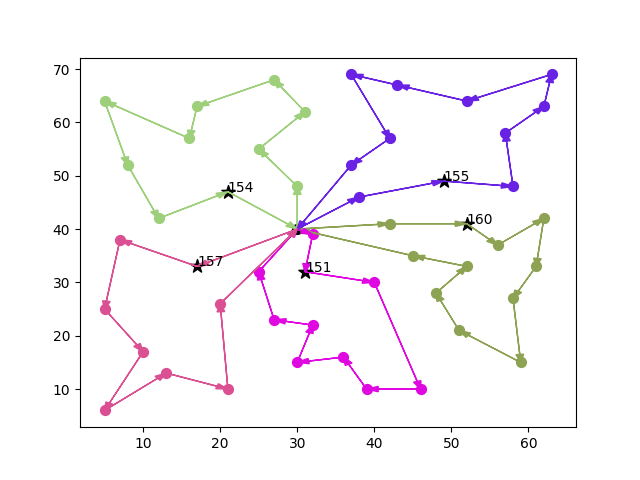}
\end{subfigure}%
\begin{subfigure}{.5\textwidth}
  \centering
  \includegraphics[width=1\linewidth]{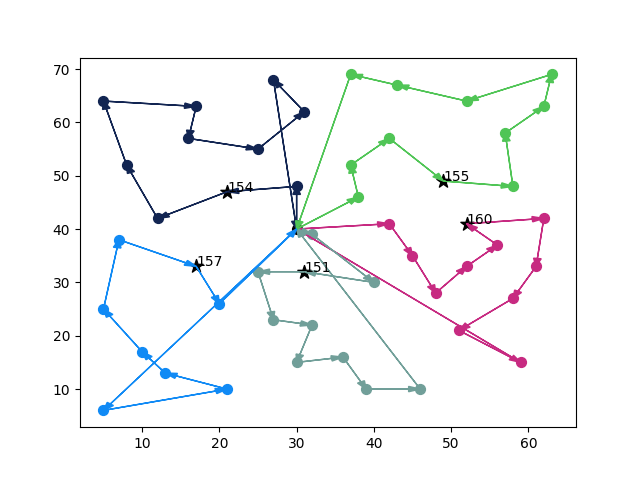}
\end{subfigure}
\caption{Paths obtained by the OR-Tools routing algorithm (Left) and \ac{qubo} routing algorithm (Right) on the clusters generated by the K-Medoids algorithm for the \textit{CMT01} data-set}
\label{fig:or_tools_qubo_cmt01}
\end{figure}

\begin{figure}
\centering
\begin{subfigure}{.5\textwidth}
  \centering
  \includegraphics[width=1\linewidth]{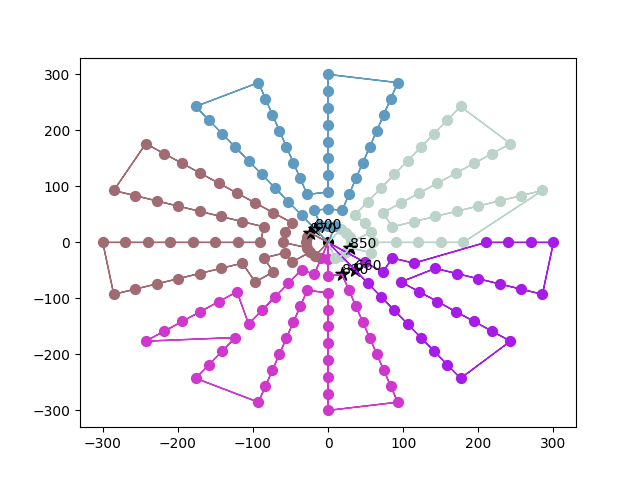}
\end{subfigure}%
\begin{subfigure}{.5\textwidth}
  \centering
  \includegraphics[width=1\linewidth]{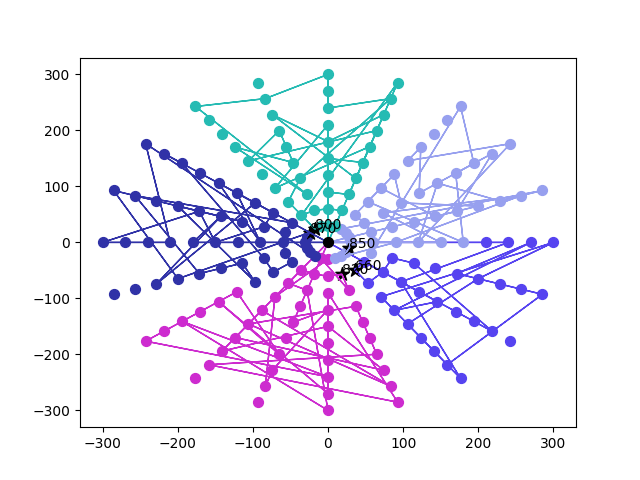}
\end{subfigure}
\caption{Paths obtained by the OR-Tools routing algorithm (Left) and \ac{qubo} routing algorithm (Right) on the clusters generated by the K-Medoids algorithm for the \textit{Golden\_05} data-set}
\label{fig:or_tools_qubo_golden05}
\end{figure}

Images \ref{fig:or_tools_qubo_cmt01} and \ref{fig:or_tools_qubo_golden05} show the paths obtained by both routing algorithms for data-sets \textit{CMT01} and \textit{Golden\_05} \cite{golden} respectively, we can see that the paths obtained by the OR-Tools routing algorithm are more optimal than the ones obtained by the \ac{qubo} routing algorithm.

\begin{figure}[!htbp]
	\centering
	\includegraphics[width=0.5\columnwidth]{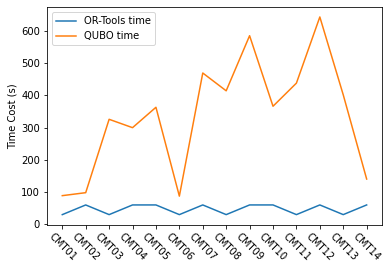}
	\caption{Running time of the OR-Tools and \ac{qubo} routing algorithms on \textit{CMT 01-14} data-sets}
	\label{fig:routing_time}
\end{figure}

The OR-Tools routing algorithm also performs better than the \ac{qubo} equations in the time taken to obtain the results, image \ref{fig:routing_time} shows the time cost of each algorithm on some of the data-sets tested, these results are expected since the OR-Tools is designed to run on a classical computer and the \ac{qubo} equations are designed to run on a quantum annealer, the experiments were performed on a classical computed using the Simulated Annealing Sampler which increases the time cost for the \ac{qubo} equations.

\FloatBarrier

\subsection{QUBO CVRP Algorithm}
\label{section:qubo_solver_analysis}

The \ac{cvrp} \ac{qubo} algorithm explained in chapter \ref{chapter:qubo_solver} can only be used in toy problems due to the high number of constraints and variables used in the \ac{qubo} equation, for the smallest \textit{CMT} data-set (\textit{CMT01}) with 50 customers and 5 vehicles the resulting \ac{qubo} has over 10,000,000 variables, most of these come from the slack variables needed in the closed loop sub-tour elimination constraint \ref{eq:vrp_c6}, with only $\pm$ 14,500 coming from the main equation \ref{eq:vrp_qubo} and the capacity constraint \ref{eq:vrp_c5}, this makes it unfeasible to use on any real data-set.
\\Even on toy data-sets the performance of the \ac{cvrp} \ac{qubo} solver is dismal, it is unable to find a good solution on a data-set with only 6 customers and 2 vehicles, and it generates a \ac{qubo} with 457 variables, 72 are the main decision variables of the problem, 200 come from the capacity constraint and 129 from the closed loop sub-tour elimination constraint, the results of the experiment are shown below:

\begin{verbatim}
  Running...
Same city penalty: 9999999999
Energy of the solution:  232.0
Time elapsed: 370.1921229362488
[[(0, 5), (3, 3), (3, 7), (5, 3)], [(0, 4), (0, 7), (1, 7), (4, 1)], 
 [(0, 0), (0, 6), (2, 7), (4, 2), (6, 7)]]
Total distance of all routes: 20000000240.77898

Checking constraints:
   FAIL -> customer 3 is visited 2 times
   FAIL -> Vehicle 2 never arrives at 4 but leaves - (4, 2)
   FAIL -> Vehicle 0 travels between the same cities - (3, 3)
   FAIL -> Vehicle 2 travels between the same cities - (0, 0)
   FAIL -> Vehicle 2 ends at depot 0 - (0, 0)
   FAIL -> Vehicle 2 has capacity 50 and customer load 90
   FAIL -> Vehicle 1 leaves depot 0, 2 times - [(0, 4), (0, 7), (1, 7), (4, 1)]
   FAIL -> Vehicle 2 leaves depot 0, 2 times - [(0, 0), (0, 6), (2, 7), (4, 2), (6, 7)]
\end{verbatim}

\begin{figure}[!htbp]
	\centering
	\includegraphics[width=0.8\columnwidth]{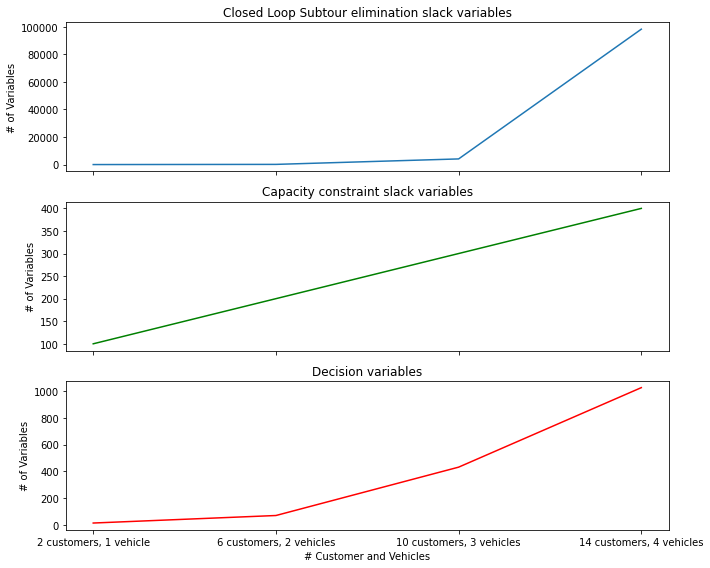}
	\caption{Number of variables generated by the full \ac{qubo} solver for different number of customers and vehicles}
	\label{fig:full_qubo_vars}
\end{figure}

Image \ref{fig:full_qubo_vars} shows how the number of variables used by the \ac{qubo} grows as we increase the number of customer nodes and vehicles, we can see that the slack variables used by the capacity constraint increase linearly with the number of vehicles used, the main decision variables and the slack variables used by the closed loop sub-tour elimination both grow exponentially with the number of customers but the main decision variables grow at a much slower rate, the number of main decision variables is also dependent on the number of vehicles while the number of closed loop sub-tour elimination slack variables scales only with the number of customers.

\FloatBarrier

\subsection{Findings}
\label{section:analysis_conclusions}

From the experiments performed above in section \ref{subsection:clustering_phase_analysis} we conclude that the best algorithm for the clustering phase of the hybrid algorithm is the \ac{qubo} clustering algorithm with the multiplier $M_1$ set to 50,000 and multiplier $M_2$ set to 20, although it offers worse time performance than the K-Medoids algorithm the clusters obtained are usually better, the K-Medoids does perform better when the data points have a clear pattern, but typically real world data does not appear with a clear pattern, thus the \ac{qubo} clustering algorithm is better suited for real problems.
\\For the routing phase, we can see in section \ref{subsection:routing_phase_analysis} how the OR-Tools routing algorithm outperforms the \ac{qubo} routing algorithm both in running time and quality of the solution obtained, thus it is the best option for the Hybrid algorithm.
\\Section \ref{section:qubo_solver_analysis} details the performance of the full \ac{qubo} solver, although we have not run the same experiment using both the full \ac{qubo} solver and the hybrid algorithm, it is clear from the experiments run on each of the algorithms independently that the Hybrid algorithm vastly outperforms the full \ac{qubo} solver.
\\Table \ref{tab:bks_results} shows a comparison between the best known solutions for the \textit{CMT 01-05} data-set and the results obtained by the hybrid algorithm using the \ac{qubo} clustering algorithm and the OR-Tools routing algorithm, the best known solutions were obtained from \textit{Solution of capacitated vehicle routing problem with invasive weed and hybrid algorithms} \cite{cmt_bks}.

\begin{table}[!htbp]
\centering

\begin{tabular}{|c|c|c|c|}
\hline
\textbf{Problem Instances} & \textbf{BKS*} & \textbf{Hybrid} & \textbf{\% Difference with BKS} \\ \hline
\textbf{CMT1(50)} & 524.61 & 537.37 & 2.43 \\ \hline
\textbf{CMT2(75)} & 835.26 & 917.95 & 9.90 \\ \hline
\textbf{CMT3(100)} & 826.14 & 933.94 & 13.05 \\ \hline
\textbf{CMT4(150)} & 1028.4 & 1161.26 & 12.92 \\ \hline
\textbf{CMT5(199)} & 1291.3 & 1344.5 & 4.12 \\ \hline
\multicolumn{1}{|l}{* Best Known Solution} & \multicolumn{1}{l}{} & \multicolumn{1}{l}{} & \multicolumn{1}{l|}{} \\ \hline
\end{tabular}%

\caption{Comparison between the best known solutions and the results of the hybrid algorithm for the \textit{CMT 01-05} data-sets}
\label{tab:bks_results}
\end{table}

\section{Conclusions}
\label{chapter:conclusions}

In this paper we attempted to solve the \ac{vrp} problem with capacity constraints, single-depot and a homogeneous fleet of vehicles, using both quantum and classical approaches.
\\The work aimed to understand the potential for a real business use case scenario, focusing on the Last Mile delivery; To this end data-sets with a relatively large number of customers and vehicles were used in the experimentation.
\\
\\Understanding the current state of development of the related technologies, we proposed a strategy for a hybrid approach, combining and comparing both Classical and Quantum algorithms.
\\The hybrid algorithm models the \ac{vrp} problem using a 2-phase approach: clustering/grouping customers and route optimization, this approach is known as a cluster-first, route-second algorithm; For each of the two phases we developed both a quantum and a classical algorithm in order to compare them and determine the most effective combination. 
\\We also developed a fully quantum algorithm to solve the \ac{vrp} problem by modeling the problem as a QUBO equation subject to multiple constraints.

\begin{table}[!hbt]
\centering

\begin{tabular}{|c|cc|}
\hline
 & \multicolumn{1}{c|}{\textbf{Clustering}} & \textbf{Routing} \\ \hline
\textbf{Classical} & \multicolumn{1}{c|}{K-Medoids} & Combinatorial Optimization \\ \hline
\multirow{2}{*}{\textbf{Quantum}} & \multicolumn{1}{c|}{QUBO Clustering} & QUBO Routing \\ \cline{2-3} 
 & \multicolumn{2}{c|}{QUBO CVRP Algorithm} \\ \hline
\end{tabular}%

\end{table}

The evaluation of the quantum algorithms on real quantum hardware has an increased complexity due to the difficulty of embedding the problems on the topology of the available hardware, this is known as the minor embedding problem, currently the biggest quantum annealer offered by Amazon Braket (D-Wave Advantage System 6.1) has a limit of around 145 fully connected variables.
\\Considering the size of the QUBO equations generated by the developed quantum algorithms and the current condition of the quantum annealers, the additional complexity of the embedding problem decreases the viability of applying the quantum algorithms for the last mile delivery problem at scale.
\\
\\Classical algorithms typically perform better than their quantum counterparts, though this is not a totally fair comparison since on one hand we have fine-tuned algorithms running on classical computing hardware, and on the other hand we have \ac{qubo} formulations running on quantum annealers in the \ac{nisq} era hardware, both technologies are in wildly different edges of the technology maturity ladder.
\\Despite this disadvantageous situation, we found that under certain circumstances, the quantum clustering algorithm presents an advantage over its classical counterpart, mainly in the scenarios where the "clusterability" rate of the data is lower.
\\The K-Medoids clustering algorithm performs better when the data can be clearly separated into clusters, however when the data presents a lower rate of "clusterability", or a higher degree of randomness, the quantum clustering approach delivers better results.
\\We consider this an outstanding finding as it sets the basis for future research into developing quantum algorithms for the constrained clustering problem.
\\Our findings in the evaluated scenarios lead us to infer that a bigger advantage may be achievable in future versions of quantum hardware, where more qubits and a more interconnected topology may provide better results at larger scales.
\\
\\The \ac{qubo} formulation of the \ac{cvrp} problem given in this paper does not offer good results when compared to the hybrid algorithm due to the increased complexity of the \ac{qubo} formulation, thus it does not present a realistic alternative.
\\For the business analysis, we identified the potential for a cost-effective relationship between the cost of running a quantum algorithm and the quality of the results obtained, specially when the data shows a higher degree of randomness, as is usually the case with real customer location data, this demonstrates a theoretical advantage for the quantum computing approach when applied to the constrained clustering problem.

\newpage

\bibliography{references.bib}  

%

\end{document}